\documentclass[journal,preprint]{vgtc}                
\ifpdf
  \pdfoutput=1\relax                   
  \usepackage{graphicx}                
  \DeclareGraphicsExtensions{.pdf,.png,.jpg,.jpeg} 
\else
  \usepackage{graphicx}                
  \DeclareGraphicsExtensions{.eps}     
\fi%

\vgtcinsertpkg

\graphicspath{{finalimages/}{./}} 

\usepackage{microtype}          
\PassOptionsToPackage{warn}{textcomp}  
\usepackage{textcomp}           
\usepackage{mathptmx}           
\usepackage{times}              
\usepackage{cite}               

\usepackage{enumitem}
\usepackage{subcaption}         

\usepackage{xspace}
\usepackage{float}
\usepackage[american]{babel}
\usepackage{hyphenat}
\usepackage{setspace,lipsum}
\usepackage{color}

\newcommand{\klpn}{\textit{Klebsiella pneumoniae}\xspace}
\newcommand{\klpnShort}{\textit{K. pneumoniae}\xspace} %

\newcommand{\epicurve}{\textit{Epidemic Curve View}\xspace} %

\newcommand{\storyline}{\textit{Transmission Pathway View}\xspace} %

\newcommand{\contactview}{\textit{Contact Network View}\xspace}

\newcommand{\linelist}{\textit{Patient Timeline View}\xspace}

\newcommand{\ch}[1]{\protect{\textcolor{black}{#1}}}

\vgtccategory{Research}
\vgtcpapertype{application/design study}

\title{In Search of Patient Zero:\\
Visual Analytics of Pathogen Transmission Pathways in Hospitals}

\author{T.\ Baumgartl$^*$, M.\ Petzold$^*$, M.\ Wunderlich, M.\ H\"{o}hn, D.\ Archambault, M.\ Lieser, A.\ Dalpke, S.\ Scheithauer,\\ M.\ Marschollek, V.\ M.\ Eichel, N.\ T.\ Mutters, \textit{HiGHmed Consortium}, and T.\ von Landesberger}
\authorfooter{
\item $^*$ T. Baumgartl and M. Petzold are shared first authors. 
\item
 T.\ Baumgartl, M.\ Wunderlich, M.\  H\"{o}hn and T.\ von Landesberger are with TU Darmstadt, Darmstadt, Germany. T. von Landesberger is now with Universit\"{a}t Rostock, Germany. E-mail: office@gris.tu-darmstadt.de.
\item
 D.\ Archambault is with Swansea University, Swansea, United Kingdom E-mail: D.W.Archambault@Swansea.ac.uk.
\item M.\ Lieser, M.\ Petzold, V.\ M.\ Eichel, N.\ T.\ Mutters are with University Hospital Heidelberg, Heidelberg, Germany, E-mail: \{first.last\}@med.uni-heidelberg.de.   
\item A.\ Dalpke is with TU Dresden, Dresden, Germany.  E-mail: alexander.dalpke@ukdd.de.
\item S.\ Scheithauer is with University Medicine G\"{o}ttingen, Universit\"{a}t G\"{o}ttingen, Germany. E-mail: simone.scheithauer@med.uni-goettingen.de.
\item M.\ Marschollek is with L. Reichertz Institute for Medical Informatics, Hannover, Germany. E-mail: Marschollek.Michael@mh-hannover.de.
\item This paper is contributed by HiGHmed infection control study group.
}

\shortauthortitle{Baumgartl \MakeLowercase{\textit{et al.}}: Exploration of Pathogen Transmission Pathways}

\abstract{Pathogen outbreaks (i.e., outbreaks of bacteria and viruses) in hospitals can cause high mortality rates and increase costs for hospitals significantly. An outbreak is generally noticed when the number of infected patients rises above an endemic level \ch{or the usual prevalence of a pathogen in a defined population.} Reconstructing transmission pathways back to the source of an outbreak -- the \textit{patient zero} or \textit{index patient} -- requires the analysis of microbiological data and patient contacts. This \ch{is often} manually \ch{completed} by infection control experts. We present a novel visual analytics approach to support the analysis of transmission pathways, patient contacts, the progression of the outbreak, and patient timelines during hospitalization. Infection control experts applied our solution to a real outbreak of \klpn in a large German hospital. Using our system, \ch{our experts} were able to scale the analysis of transmission pathways to longer time intervals (i.e., several years of data instead of days) and across a larger number of wards. Also, the system is able to reduce the analysis time from days to hours. \ch{In our final study, feedback from twenty-five experts} from seven German hospitals provides evidence that our solution brings significant benefits for analyzing outbreaks. It is also applicable to COVID-19 hospital-associated transmissions. 
} 


 \CCScatlist{ 
 \CCScat{H.5.2}{User Interfaces}%
{User-centered design};
 \CCScat{I.2.1}{Applications and Expert Systems}{Medicine and science};
  \CCScat{E.1}{Data Structures}{Graphs and networks}
}

\teaser{
  \centering
 \includegraphics[trim=0mm 0mm 0mm 0mm, clip,width=1\linewidth]{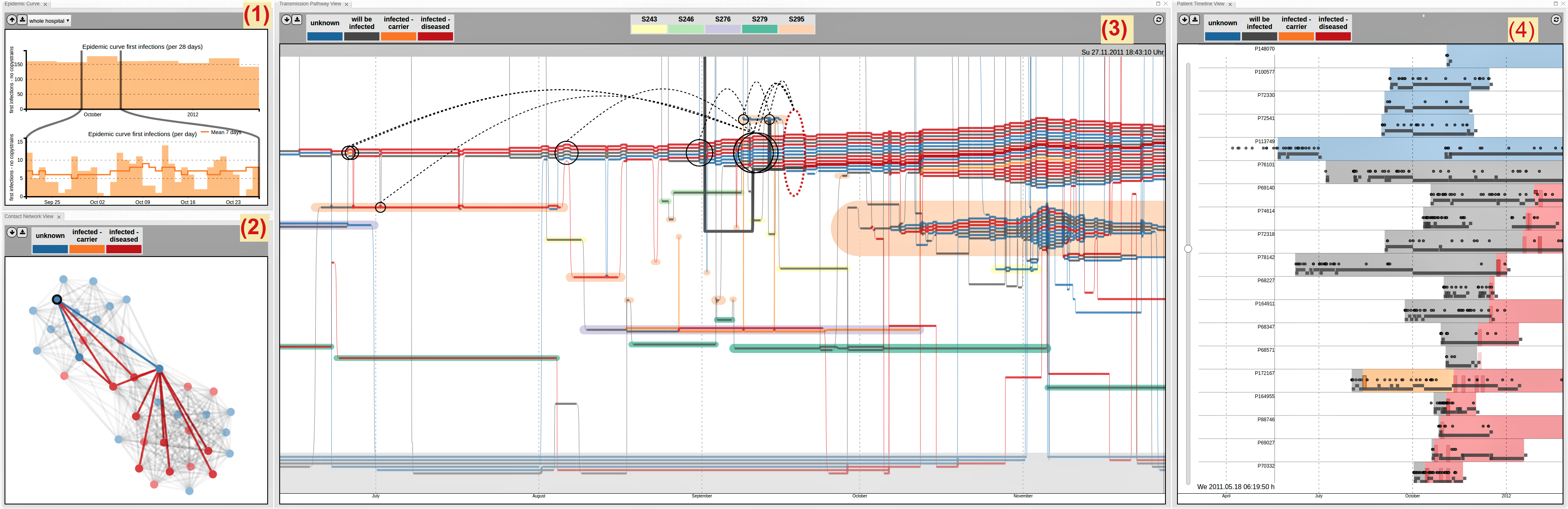}
  \caption{  \ch{Transmission pathway tracing interface. 1.\ The \epicurve shows the number of infected patients. 2.\ The \contactview shows patient contacts as a network.  3.\ The \storyline shows contacts and infection status. Tracing interaction shows potential infection transmission events.  4.\ Patient event details are shown in \linelist.}}
  \label{fig:teaser}
}

\begin{document}


\firstsection{Introduction}

\maketitle









 Pathogen transmissions are an acute problem in hospitals around the world~\cite{MDRO,engler2018hospital}. The transmission of pathogens, such as bacteria and viruses, can endanger the lives of patients, since they represent a very vulnerable group of persons. Pathogen infections in hospitals can be transmitted via patient-to-patient contact in the same room or ward~\cite{donker2012hospital}. Transmissions are generally difficult to detect since patients may be infected, but show no clinical symptoms. 
 These \textit{carriers} are invisible sources for potential transmissions. Screenings can detect carriers, but regular screening of all patients is very costly \ch{and ineffective. Thus, only high risk} patients are usually screened when entering the hospital or after transfers within the hospital. 
 
 
One initial infected patient, a \textit{patient zero} or \textit{index patient}, may transmit the pathogen to other patients \cite{engler2018hospital}. These patients may change wards and infect further patients. 
The endemic level represents the usual prevalence of a pathogen in a defined population. \ch{When the number of infected patients rises above a certain endemic level, and a spatio-temporal context is observed, it is called \textit{cluster} or \textit{outbreak}. The endemic level is characterized by parameters such as the type of patients and the location within the hospital (underlying diseases), as well as the pathogen specifics: e.g., the mode of transmission, seasonality of occurrence, the resistance potential and infectiousness. As automatic outbreak detection is still an ongoing research problem~\cite{enki2016comparison,zacher2019supervised}, usually, the endemic level is set manually by comparing numbers of newly infected patients within a certain time frame with the numbers of infected patients in a previously recorded time frame. As not all patients can be regularly tested or screened, depending on the pathogen, it can take days, weeks or months before an outbreak is detected~\cite{haarbrandt2018highmed}.} 
Therefore, the monitoring of infectious persons and determining their transmission pathways is the primary goal of Infection Control experts, e.g., hygienists, clinicians, hygiene experts, in order to intervene in time and prevent further pathogen spreading. 

Once an outbreak has been detected (Task 1), the infection control experts need to trace the infection back to its source in order to determine if the patients are connected and belong to an outbreak. This means the identification of all \ch{potentially} infected patients back to the \textit{patient zero}, i.e., the overall source of the outbreak (Task 2). The experts need to reconstruct these \textit{transmission pathways} --  by whom, when and where did a transmission occur. This information is used to find the origin of the infectious agent: whether it is nosocomial -- \ch{hospital-intern}. 
Furthermore, patient localization (which rooms/wards -- Task 3) and outbreak duration (begin/end -- Task 4) need to be determined. The transmission pathway is then used to identify putative colonized or infected patients that are yet unknown and thus require testing (Task 5). Intervention procedures can then be implemented by isolating or cohorting affected patients in separate rooms\ch{, special disinfecting processes, increasing hand hygiene, screening and teaching
staff affected: infected and exposed.} 

The identification of transmission pathways is challenging because it can occur over several contacts, involve several hospital wards and various time spans (days to years). 
Tracing requires integrating both the spatio-temporal patient information and microbiological test results. Outbreak pathway reconstruction needs to be fast and precise: each potential patient and their contacts need to be detected to prevent further disease spreading. Often there is uncertainty about the patient status at the time of a contact, as screening often detects an infection or colonization \ch{only} at a later date. Thus, also potential infections need to be considered.   

Currently, transmission pathway reconstruction is a time-consuming and potentially error-prone manual process. It may take days to weeks, using current hospital systems~\cite{haarbrandt2018highmed}. Visual analytics systems have the potential to support this analysis process by saving analyst time~\cite{Chen2014SaveTime}. However, current solutions focus mainly on the disease evolution at a population level (see \autoref{sc:related}).

In cooperation with Infection Control experts from four German Hospitals and Infection Research Institutes, we developed a novel visual analytics system for the exploration of disease outbreaks. We used iterative, user-centered design within a common project \href{https://www.highmed.org/about/use-cases/infection-control}{HiGHmed} over two years. The system offers several specialized views for the exploration of outbreaks and pathogen transmissions. The core contribution  is a novel view for contract tracing that was inspired by the well known storyline visualization~\cite{reda2011visualizing,liu2013storyflow,SoftwareEvolutionStorylines}. The visual design, layout and interactions enable to explore contacts as well as to automatically determine and highlight patients and their contacts that could transmit pathogenic organisms.

Our approach was applied to a real outbreak of \klpn in a large German hospital. The experts were able to effectively reconstruct the transmission pathway back to the \textit{patient zero} in a faster and in a more comprehensive way when compared to existing methods. \ch{In our final qualitative study, we} gathered feedback from twenty-five experts located in seven German hospitals. The results indicated significant added value when tracing transmission and analyzing outbreaks in hospitals.

\section{Related Work}
\label{sc:related}

\paragraph{Visual Analytics of Health Data.}
Visual Analysis of healthcare data is an active area of research.  Event-based visual analytics approaches for health record data has been \ch{a main  focus of this area}~\cite{lifeflowSurvey,caballero2017visual,hakone2016proact,rind2017visual,Schmidt2019,monroe2013temporal,EventThread}.  
Disease surveillance and epidemiology has also been of interest~\cite{preim2020survey}. Visual analytics approaches for epidemiologists focus on the spatio-temporal evolution of a disease at a population level--how many people will be infected, in which geographic area(s), and the speed of the spread~\cite{maciejewski2011pandemic,bryan2015integrating,yanez2017pandemcap}.
These visual analytics approaches target disease spreading over a large population at a macroscopic level. They visualize the number of people infected over time, but detailed views about individual patient contacts are not the focus.  Machine learning methods working with visual analytics have been proposed~\cite{choi2016retain,kwon2018retainvis}\ch{, and more specifically for disease progression pathways~\cite{kwon20}}.  They have been applied to the problem of infection control~\cite{EuroVA2020}, but in this closely related work the transmission pathway was not reconstructed.  Machine learning is effective, but may not be feasible for rarely occurring pathogens, such as \klpnShort, where the numbers of items in the trained classes of (non)-infection are very biased. \ch{Thus,} transmission pathway reconstruction and the identification of potentially infected patients requires interactive exploration.

\paragraph{Disease Spread and Dynamic Networks.}
As \ch{many} pathogens are transmitted over contacts preferentially, disease spreading is often simulated as dynamic processes over contact networks. A number of approaches have focused on the population level:  the number of infected patients through line charts~\cite{NEMOSimVis,donker2012hospital,liang2010simulation,sahneh2017gemfsim}. They use the contact networks in the simulation, but only provide macroscopic views of disease progression.
The simulations often result in a collection of thousands of dynamic graphs with very few existing approaches for this data~\cite{liu2019aggregated,beck2017taxonomy}.
Manynets~\cite{freire2010manynets}, GraphLandscape~\cite{kennedy2017graph} and SOM-based clustering~\cite{von2009visual} extract graph properties and compare many static graphs and these methods could be adapted and applied.  However, two very different graphs can have the same properties~\cite{chen2018same}. \ch{Network piling} approaches~\cite{bach2015small,vorgogias19} allow the comparison of several static graphs in detail but do not scale well to larger graphs and have not been adapted to compare multiple dynamic graphs. Recently, a simulation and visualization of disease spreading over contact networks in hospitals has been presented~\cite{wunderlich2019visual}. It shows the disease and transmission probability on individual level, but takes only a static network as input. This simplification is done for simulation purposes.  However, in real cases, patients move across wards over time. Both visualization and simulation should take into account the network dynamics over time. In sum, however, the simulations focus only on disease spreading prediction and not on the reconstruction of pathogen transmission pathways.

\paragraph{Visualization of Contact Dynamics.} The reconstruction of pathogen transmission pathways requires analyzing patient contacts and patient infection status over time.  
A number of methods exist for visualizing a single dynamic graph over time in a scalable way~\cite{von2015visual,van2015reducing,bach2013graphdiaries,alluvial}. These methods use timeslices as a basis for the visual analytics process. The uniformity of timeslices poses a problem as the distribution of transmission events is unequal over time.  Event-based methods~\cite{monroe2013temporal,qi2018visual} for the visualization of dynamic networks~\cite{Lee19Plaid,DynamicGraphsWithoutTimeslices,eventBasedDG,NonuniformTimeslicing} are more applicable. However, they have not been designed for outbreak networks visualization in a way that fulfills the needs of the infection control experts.

The visualization of contact network dynamics for hospital data could be supported by dynamic set visualization and storyline approaches. Set visualization methods~\cite{alsallakh2016state} for the visualization of set dynamics~\cite{von2012visual,nguyen2016timesets} are able to show the set sizes and the number of changes between sets over time. \ch{Methods for visualizing long time series~\cite{timecurves} and text visualizations~\cite{textflow,trainsofthought} have similar representations.}  Showing the group membership over time on individual level has \ch{also} motivated the storyline research.
Storyline visualizations~\cite{padia2018yarn,padia2019system,tang2018istoryline,SoftwareEvolutionStorylines,tanahashi2012design,liu2013storyflow} devote one dimension, usually the x-axis, to time and encode each individual character in a story as a line.  Lines that are placed close together to indicate the characters share a scene; when the lines separate, the scene ends. For disease spreading, this encoding can be used to show contacts in the same ward and forms a key part of our approach. Research in storyline visualization has focused on optimizing the compactness of storyline visualizations (either automatic or users-assisted)~\cite{arendt2017matters,froschl2017minimizing,liu2013storyflow,padia2018yarn,padia2019system,qiang2016storycake,silvia2015storyline,silvia2016variantflow,tanahashi2012design,tang2018istoryline}, reducing crossings~\cite{gronemann2016crossing,kostitsyna2015minimizing,van2016block,wolff2018computing}, plotting approaches~\cite{shrestha2013storygraph}, combining storylines with event-based methods~\cite{arendt2016too}, genealogical data~\cite{kim2010tracing}, streaming and dynamic data~\cite{tanahashi2015efficient,yagi2015layout}, and contacts between living things or actors exhibiting similar behavior~\cite{reda2011visualizing}. Reda {\it et al.}~\cite{reda2011visualizing} is the closest approach to ours, but it needs to consider all contacts in the storyline. In our work, we can reduce \ch{clutter by determining and highlighting pathogen transmitting contacts.  On the other hand, we need to show more specific locations inside the hospital or if the patient is discharged.}

\section{Background -- Tasks and Data}
\label{sc:datatasks}

\ch{Over the course of one year, we investigated the data and tasks that are performed by infection control experts in four German hospitals during their procedures. The data and tasks were guided by the scope of the \href{https://www.highmed.org/about/use-cases/infection-control}{HiGHmed} project. For a deep understanding of the data, tasks and current analysis methods, we conducted a structured interview following the methodology in~\cite{von2016visualization}. We first used an online questionnaire that was answered by six infection control experts and epidemiologists. Moreover, we interviewed three hygienists to learn the current work methods by infection control experts in three hospitals and one infection control institute.  Further details on the tasks and data were assessed during the iterative design process (see \autoref{sc:designprocess}). Our system is designed to support one of their main tasks: tracing transmission pathways in outbreaks.}

\subsection{Tasks}
\label{sc:tasks}

Our work resulted in these \ch{tasks} without a strict workflow order:
\begin{itemize}[nosep]
    \item[T1] \textit{Detect Outbreak.}
    \ch{Is there an outbreak? When the number of infected patients rises above a normal level within a certain period, i.e.,  the endemic level, an \textit{outbreak} occurs. Depending on the pathogen, this endemic level can be two or more patients.  As patients may not be tested or screened generally, an outbreak is determined by manual inspection.} 
    \item[T2] \textit{Outbreak Pathway.} 
    \begin{itemize}[nosep]
        \item[T2.1] \textit{\ch{Determine transmission contacts.}}
        Did contacts occur between patients that could have led to pathogen transmission? If yes, where, when and with whom?
        \item[T2.2] \textit{Determine index patient (patient zero).} 
        \ch{Who is the source of the outbreak? Identifying the transmission pathway should lead back to the index patient, i.e., the patient zero.}   
        \item[T2.3] \textit{\ch{Distinguish between a single or multiple outbreaks.}} 
        Is the observed outbreak a single outbreak or multiple, simultaneous outbreaks of similar pathogens? \ch{Depending on the answer, there could be one or more index patients.}
    \end{itemize}
    \item[T3] \textit{Outbreak Location.}
    \begin{itemize}[nosep]
        \item[T3.1] \textit{\ch{Determine if the outbreak is hospital-associated.}}
        \ch{Did the transmissions occur within the hospital or have been infections brought in from outside the hospital?} A within-hospital transmission, i.e. a nosocomial, points to a source of infection such as a patient/staff/device. These transmissions should be avoided as patients can have lower immune response and thus may be more  vulnerable.
        \item[T3.2] \textit{\ch{Locate ward(s) with pathogen transmissions.}}
        \ch{For nosocomial transmissions, in which ward has the transmission occurred?} This location needs to be disinfected or closely monitored. 
    \end{itemize}
    \item[T4] \textit{\ch{Quantify outbreak duration.}} 
   \ch{ When did the outbreak first start? How long does it last?} Early detection mitigates the risk of larger outbreaks. Tracing back the time point of all transmissions to the initial index patient determines the total duration.
    \item[T5] \textit{\ch{Identify potentially infected patients.}} 
    \ch{Who is also potentially infected?} Not all patients can be screened in the hospital to determine their infection, and not all infected patients show symptoms. These potentially (i.e., putative) infected patients need to be identified via their contacts to infected patients.
\end{itemize}

\subsection{Data}
\label{sc:data}

The pathogen transmission analysis requires to combine two types of data 1) patient locations for determining contacts, and 2) microbiological data for identifying infection status. Due to privacy reasons, collection of this data is limited (see \autoref{sc:discussion}). The patient location is only determined by a log record of their ward, not by tracking. The infection status is only known at the time of screening or test. \ch{These data sets can span years with events recorded down to second precision. The time between two consecutive events (e.g. transfers) can be on the order of seconds to months.}

\begin{figure}
    \centering
 \includegraphics[width=\linewidth]{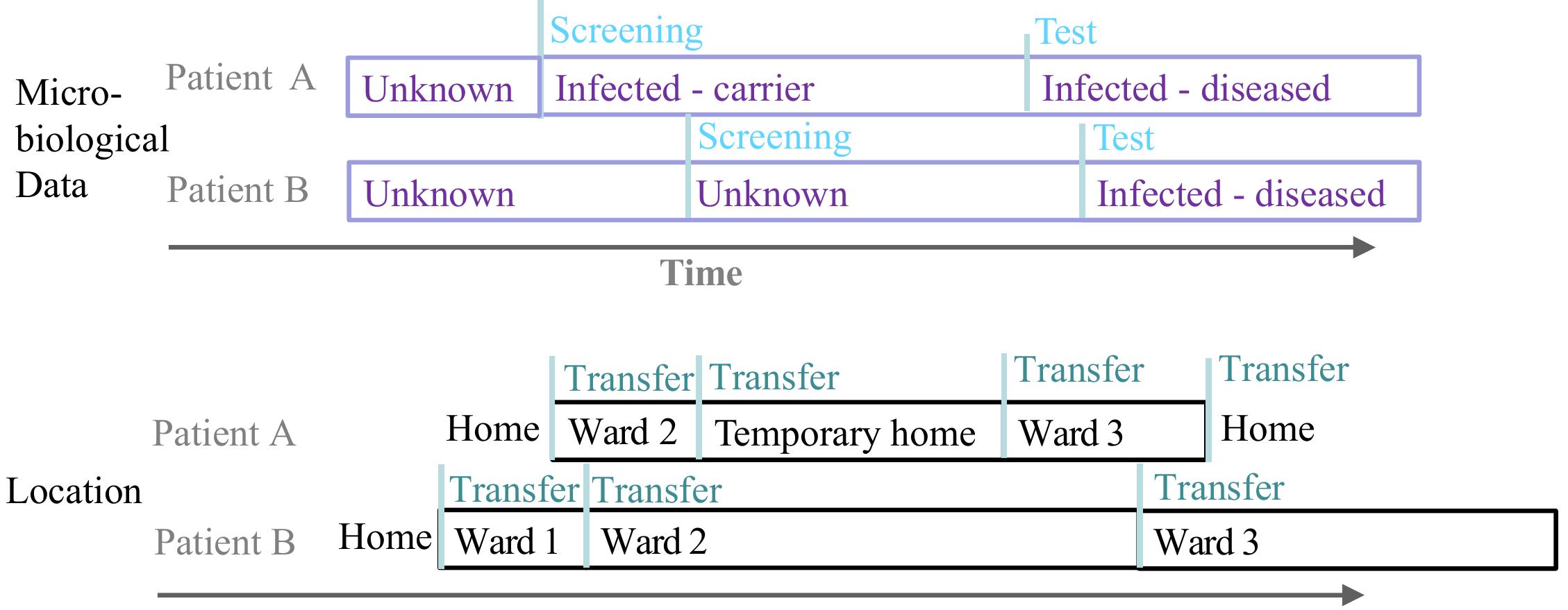}
     \caption{Data characteristics. \ch{Bottom:} Patient locations determined by transfer events. \ch{Top:} Infection status of patients over time determined by microbiological data events.}
    \label{fig:datacharacter}
\end{figure}

\paragraph{Patient Locations:} The location data consists of a list of patient transfer events $TR=\{TR_k\}$. A transfer $TR_k$ records that the patient $P_k$, was transferred to location $L_k$ at time $t^{TR}_k$ for the following reason ${type}_k$. Thus, a transfer is $TR_k=(P_k,L_k,t^{TR}_k, {type}_k)$. The ${type}_k$, determines whether the transfer was 1) between wards, 2) from home to hospital, i.e., \ch{first hospital admission, or 3) home between hospital stays, i.e., `temporary home',} or 4) was the end of a hospital stay. 
The current location of the patient is the destination of the patient's last transfer (see \autoref{fig:datacharacter}).  The time intervals between successive transfers are irregular and cannot be transformed to regular intervals without sampling or without causing scalability challenges. Even very short stays at a location can lead to important contacts in pathway reconstruction.

\paragraph{Microbiological data:} $MB$ is information about the tests and screenings of patients for pathogens. More formally, it is a set of events $MB=\{MB_j\}$. Each element $MB_j$ records which patient $P_j$, was tested or screened $s_j$, for which pathogen $\rho$, at what time $t_j^{MB}$ and the result $r_j$:  $MB_j=(P_j,s_j,t_j^{MB},\rho_j,r_j)$. 
Patients may have several microbiological data records associated with them or none at all. The analysis can take place before, during or after the hospital stay and the spacing between events is irregular.

The result $r_j$ determines patient's \textit{infection status} (see \autoref{fig:datacharacter}): 
\setlist{nolistsep}
\begin{itemize}[noitemsep]
    \item \textit{infected -- carrier}: A positive result of a screening means that the patient is colonized. Due to lack of data on recovery (see \autoref{sc:discussion}) from this moment onwards, the patient retains this status unless he/she is later identified as diseased. 
    \item \textit{infected -- diseased:}  If a patient with symptoms is tested positively on the pathogen $\rho$, he/she is in a diseased state. Due to lack of data on recovery (see \autoref{sc:discussion}), the patient retains this status from the moment of the positive test onwards.
    \item \textit{unknown:} Before the first positive screening or targeted microbiological test, the infection state is \textit{unknown}. If the patient is not tested for the pathogen $\rho$, it is unknown whether they are infected/colonized.  If a patient becomes infected at a later point in the data set, the patient is labeled \textit{``unknown -- will be infected''}.
\end{itemize}

\section{Design Process and Final Approach}
\label{sc:designprocess}



Our approach was developed through iterative design with our domain experts over two years following guidelines for effective visualization design~\cite{munzner2009nested,sedlmair2012design,kerracher2017constructing}. \ch{After the initial task and data analysis, we had regular quarterly meetings with project participants. The number of participants in the meetings varied as not all participants could attend each meeting. On average, ten of our expert participants were present in our meetings.  The expertise included: infection control, hygienists, infection control data management, epidemiologists and infection control managers. Between the meetings, we communicated per email and held interactive sessions onsite.}

\subsection{Prototype Development}
\label{sc:protoDevel}
\begin{figure}[t]
    \centering
    \begin{subfigure}{\linewidth}
      \includegraphics[width=\linewidth]{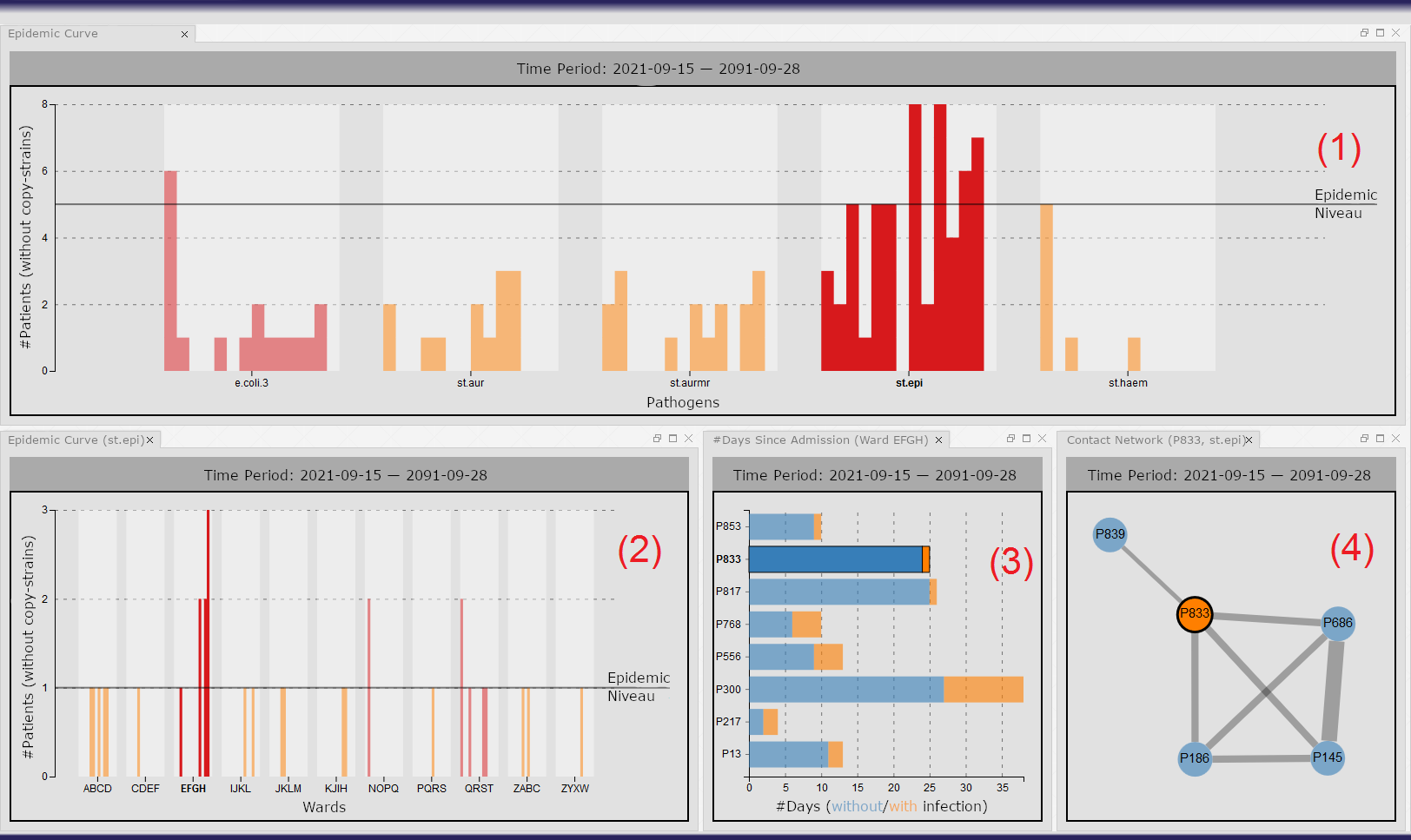}
        \label{fig:initdashboard}
        \caption{Initial visualization, see also~\cite{Dashboard}}
    \end{subfigure}
    \begin{subfigure}{\linewidth}
         \includegraphics[width=\linewidth]{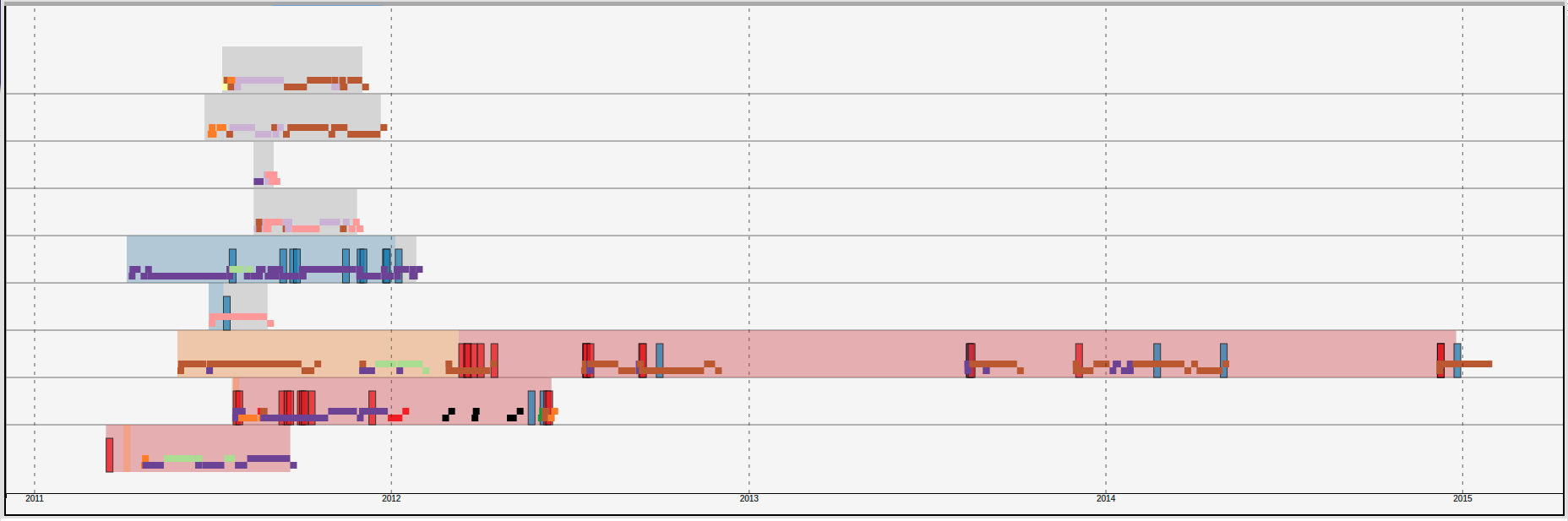}
         \caption{Patient timeline view.}
         \label{fig:linelist}
    \end{subfigure}
    \begin{subfigure}{\linewidth}
        \includegraphics[width=\linewidth]{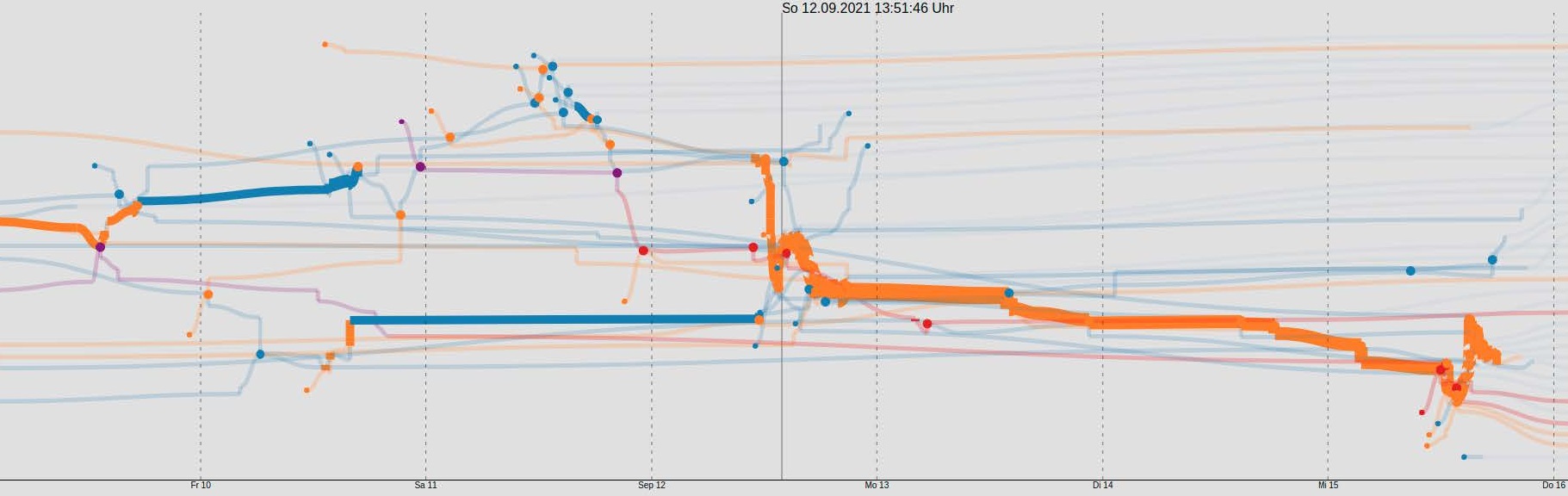}
               \caption{Initial storyline-like design}
         \label{fig:storylineold}
    \end{subfigure}
    \caption{\ch{Early prototypes. Interfaces were translated into English.}}
    \label{fig:designdevelopment}
\end{figure}

Our {\it first prototype} had several linked views~\cite{Dashboard} (see \autoref{fig:designdevelopment}a). The epidemic curve view \ch{showed} the number of infected patients over a two weeks period (see \autoref{fig:designdevelopment}a--part 1,2). It highlighted potential outbreaks with infections above the endemic level per selected pathogens. For identifying hospital-associated (nosocomial) infections, the patient stay along with infection status \ch{was} shown (see \autoref{fig:designdevelopment}a--part 3). Potential transmissions between patients were supported only by node-link diagrams showing patient contacts and current infection status (see \autoref{fig:designdevelopment}a--part 4). \ch{A total of thirteen domain experts responded to our call for feedback by filling out an online questionnaire}~\cite{Dashboard}. The prototype demonstrated a high degree of usefulness (mode 4 and 5 respectively on a five point Likert scale). The experts especially appreciated views that were new to them: a) patient stay and the infection duration and b) the contact network. They wanted these views to scale to longer periods of time, larger sets of microbiological data, and views to support pathogen transmission pathway reconstruction.

The {\it second design} included a \linelist that showed the patient stay and his/her microbiological data (see \autoref{fig:designdevelopment}b). This design leveraged the infection control expert's experience with excel-based information. Each row is one patient, and the x-axis is time. The background color shows the infection status, \ch{colored} horizontal bars show the patient location, and \ch{colored} vertical bars show the screening and test results. Our experts found that this view provided a good overview of patient location and infection status. However, the separation of patients into rows made it difficult to spot contacts leading to pathogen transmission. Sorting the rows  by the time of the first infection did not help, as a patient can have contact with several patients.

Our {\it third design} used a storyline representation of the infection data (see \autoref{fig:designdevelopment}c). Patients are lines and the x-axis is time. Lines are colored according to infection status. Contacts in this view are line bundles. Although the view shows all data needed for pathogen transmission analysis, the experts had difficulty relating to it. The layout that emphasized the movements made it difficult to follow patients and to determine transmissions. 

\subsection{Final Prototype}
The {\it final visual interface} is shown \ch{in} \autoref{fig:teaser}. This design combines the pathogen pathway reconstruction views of the user-selected pathogen needed for our tasks (see~\autoref{sc:tasks}). 
\setlist{nolistsep}
\begin{enumerate}[noitemsep,nosep,leftmargin=*,itemindent=0pt,labelwidth=0em,labelindent=0em]
    \item \textit{\epicurve}  shows the number of infected persons per day in order to support Task 1 -- outbreak detection. The infection control expert can select the total number of infections or only new ones (i.e., without copystrains). To see how it relates to the endemic level, a moving average of the user-selected time period is shown. This view can show data for the hospital (see \autoref{fig:UCEC} top) or specific wards (see \autoref{fig:UCEC} bottom).  It supports longer time periods via focus-and-context methods inspired by~\cite{walker2015timenotes}. 
    \item \textit{\contactview} shows the contacts of selected patients for determining putative infected patients (Task 5).
    \item \textit{\storyline}  supports Tasks 2--4. This view shows patient contacts and possible pathogen transmissions over time in a storyline-like view. The  extensions are layout, design, and highlighting of potential transmission events more directly. 
    \item \textit{\linelist} shows the patient location overlaid with microbiological data. It supports the visual analytics process with infection status and location information. \ch{The data encoding is unchanged since its early prototype described in Sec.~\ref{sc:protoDevel}.}
\end{enumerate}


\section{\storyline}
\label{sc:storyline}

The \storyline shows patient infection status and contacts over time and across locations (\autoref{fig:teaser} (3)). Outbreak duration, potential transmission contacts between patients, and patient locations are visible (Task 2--4).

Each line represents a patient and the x-axis encodes time (Task 4).  Each patient line starts with the earliest recorded admission to hospital and ends with the last recorded stay. Temporary home stays are also shown. This helps in detecting hospital-associated transmissions during previous stays (Task 3.1).  Patient lines pass close to each other for every potential contact (Task 2.1). The y-axis encodes patient location.  Fixed vertical positions for \ch{individual wards} (as in Baling \textit{et al.}~\cite{balint2016storyline}) is not scalable due to the larger number of wards and patients (hundreds), but our layout still aims at preserving the vertical position of wards~\cite{arendt2017matters}.  Line color conveys infection status, and background color is used to encode location information. \ch{Technical details are below.}

\subsection{Layout}
\label{sc:layout}

\begin{figure}[tbph]
    \centering
		\begin{subfigure}{\linewidth}
		  \centering
  \includegraphics[width=0.9\linewidth]{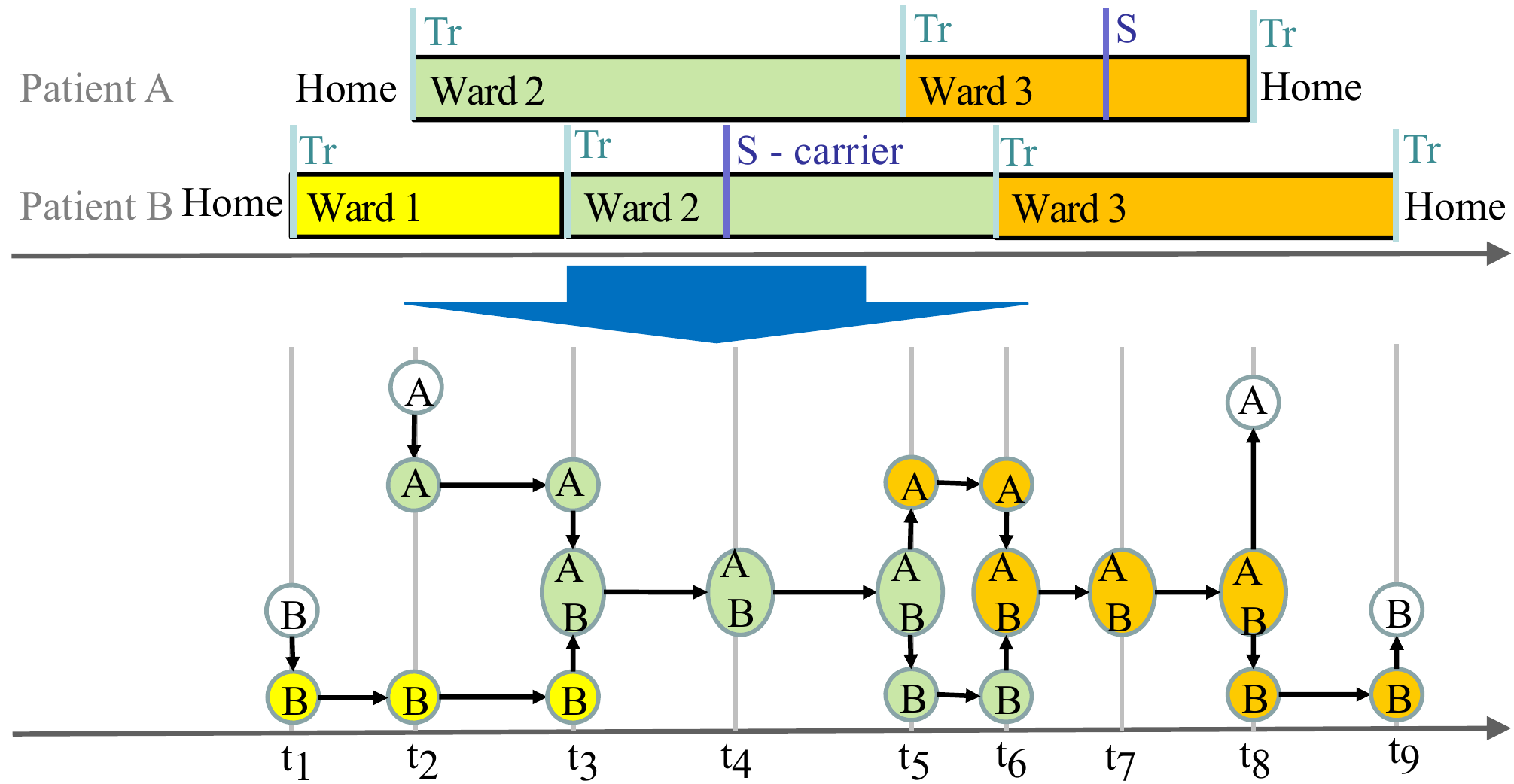}
        \caption{Pre-processing: Constructing patient contact graph}
        \label{fig:preprocess}
    \end{subfigure}
       \centering
\begin{subfigure}{0.9\linewidth}
    \centering
    \vspace{2mm}
    \textcolor{gray}{
 \setlength{\fboxsep}{0pt}%
\setlength{\fboxrule}{0.5pt}%
\fbox{\includegraphics[width=\linewidth, height=2.5cm]{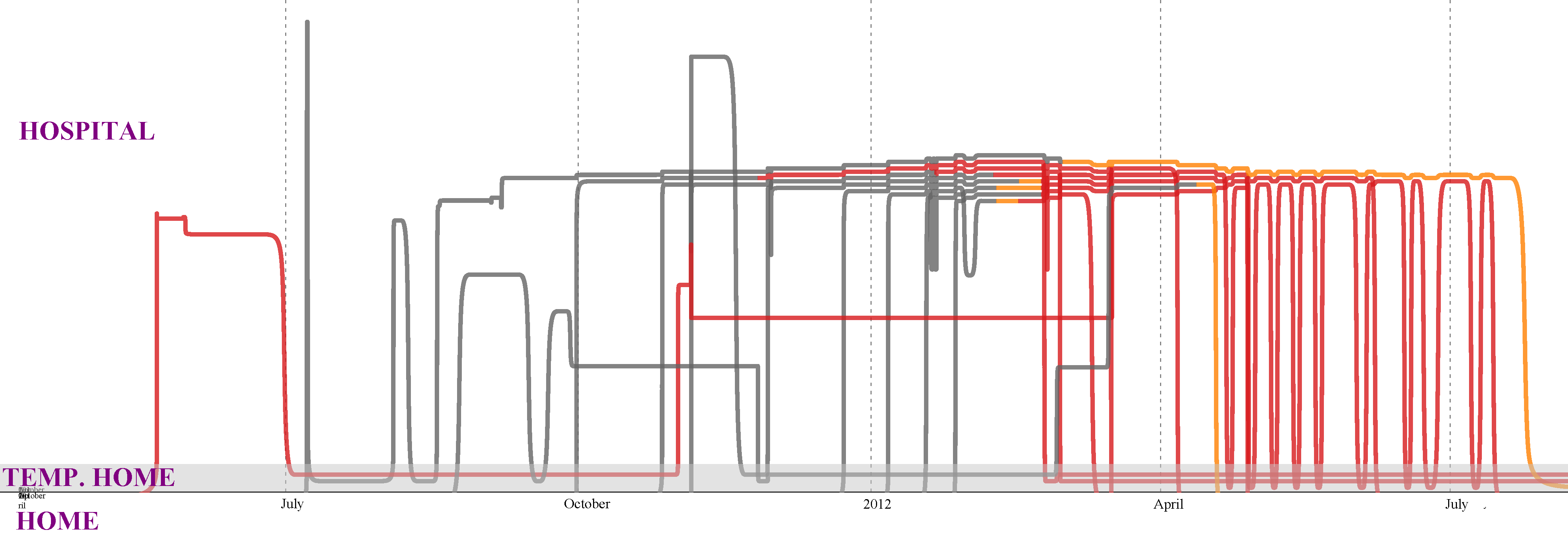}}}
    \\
     \vspace{-1mm}
    \small{\textit{This variant leads to more clutter and lower discriminability between home and 'temporary home' transfers.}}
    \vspace{2mm}
    \textcolor{gray}{
 \setlength{\fboxsep}{0pt}%
\setlength{\fboxrule}{0.5pt}%
\fbox{\includegraphics[trim= 0 0 0 0, clip,width=\linewidth, height=2.5cm]{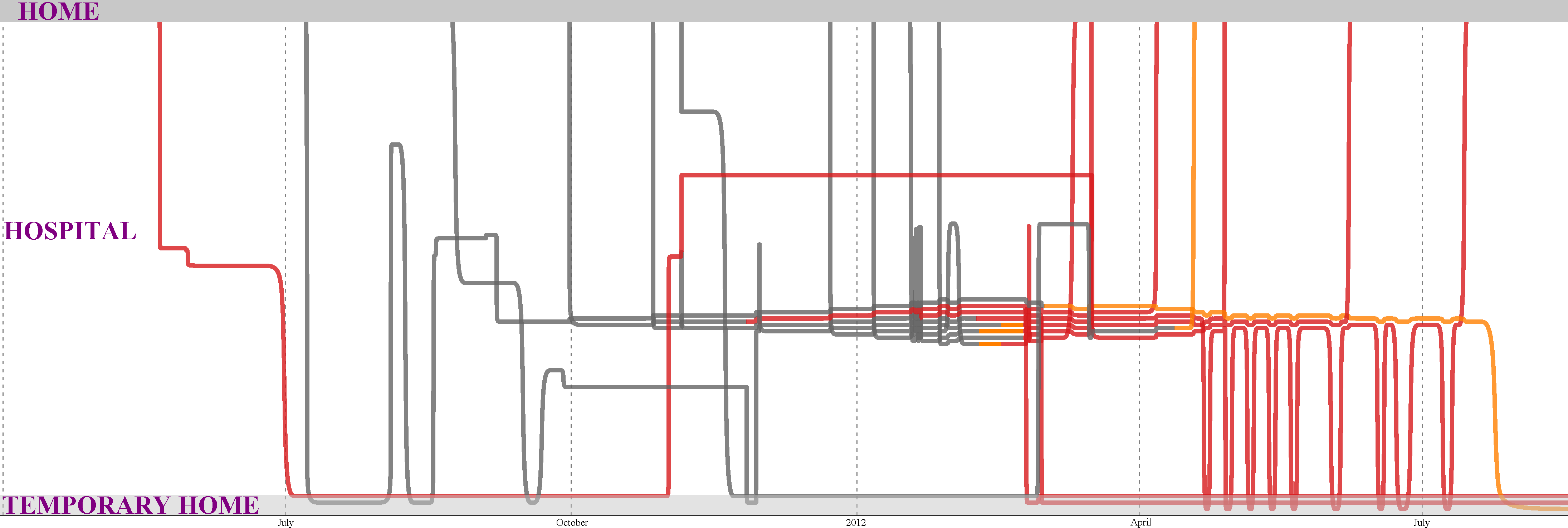}}}
    \\
     \vspace{-2mm}
    \small{\textit{The used variant distinguishes transfers from home (lines from top) and from  'temporary home'(lines from the bottom).}}
    \vspace{-2mm}
    \caption{Constraining area for home, hospital, and temporary at home.}
    \label{fig:areas}
  \end{subfigure}
    \begin{subfigure}{\linewidth}
    \vspace{2mm}
		   \centering
       \textcolor{gray}{
 \setlength{\fboxsep}{0pt}%
\setlength{\fboxrule}{0.5pt}%
\fbox{\includegraphics[width=\linewidth]{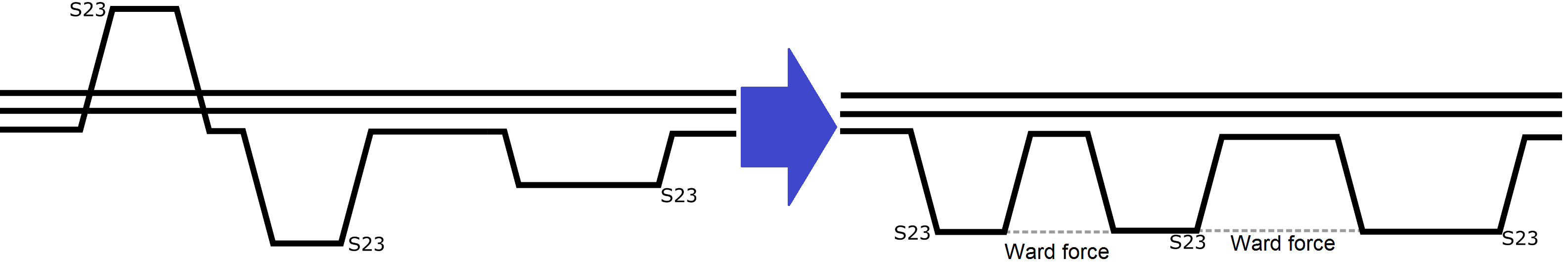}  }}
        \vspace{-3mm}
        \caption{Spring force for wards improves stability of y position.}
        \label{fig:layoutwards}
    \end{subfigure}
     \begin{subfigure}{\linewidth}
		   \centering
		   \vspace{2mm}
       
\textcolor{gray}{
 \setlength{\fboxsep}{0pt}%
\setlength{\fboxrule}{0.5pt}%
\fbox{
\includegraphics[width=1\linewidth]{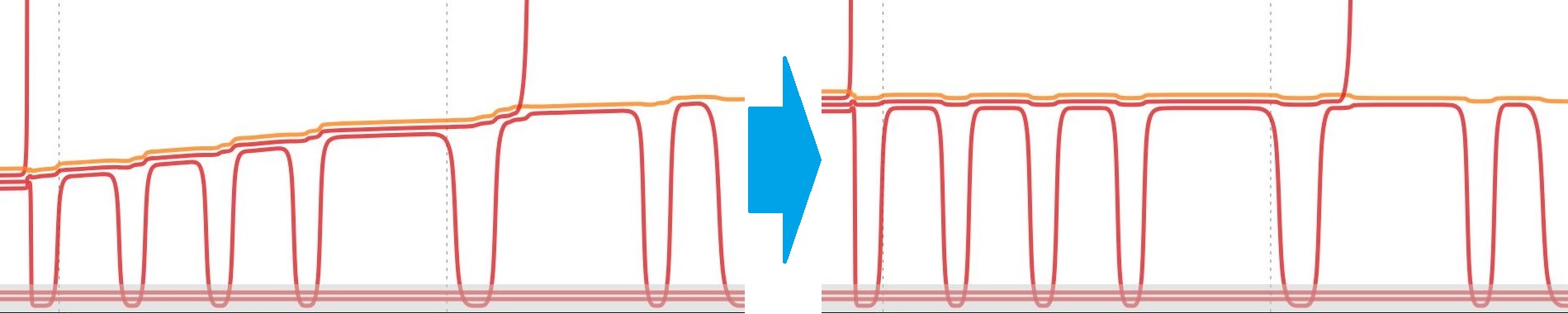}
}
}
        \caption{Line straightening is performed after each iteration}
        \label{fig:straightening}
        \end{subfigure}
       
        \begin{subfigure}{\linewidth}
				   \vspace{2mm}
		 \centering
       \setlength{\fboxsep}{0pt}%
\setlength{\fboxrule}{0.5pt}%
\textcolor{gray}{
 \setlength{\fboxsep}{0pt}%
\setlength{\fboxrule}{0.5pt}%
\fbox{\includegraphics[trim=5 50 5 25,clip,width=1\linewidth]{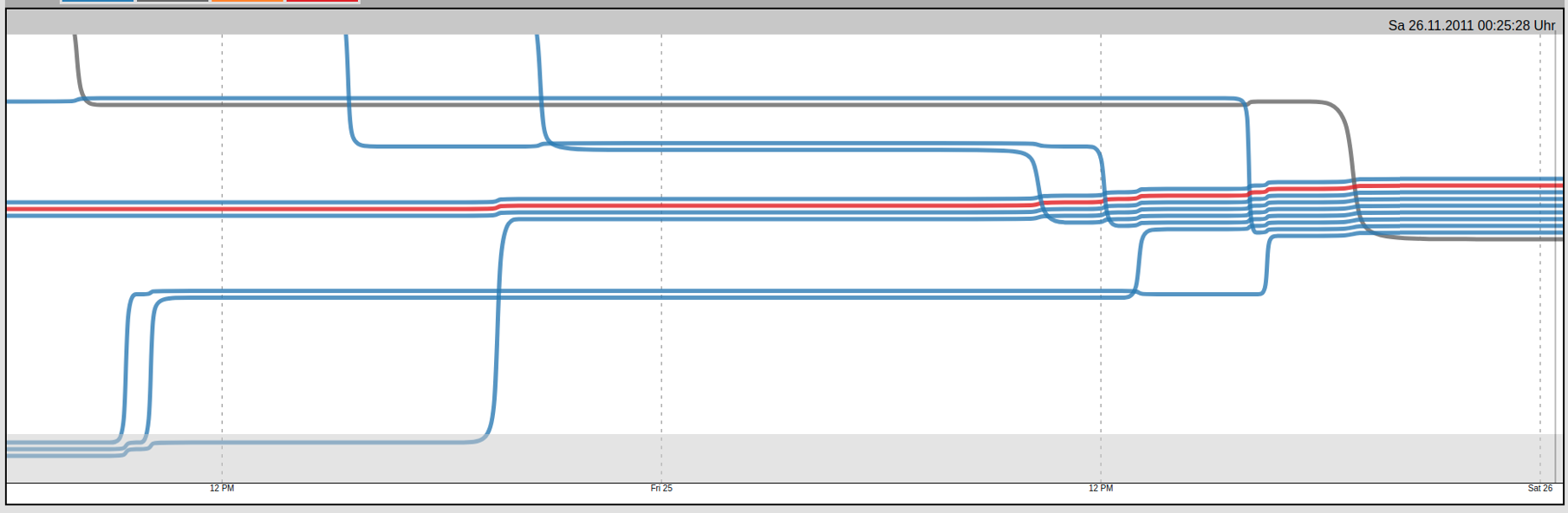}}}
       \centering
        \caption{Line order according to patient transfer order. Wiggles emphasize transfers, esp., ensure visibility of short time periods.}
        \label{fig:ordering}
    \end{subfigure}
      
         \begin{subfigure}{\linewidth}
				   \centering
        \textcolor{gray}{
 \setlength{\fboxsep}{0pt}%
\setlength{\fboxrule}{0.5pt}%
\fbox{\includegraphics[trim=0 0 0 0, clip,width=\linewidth]{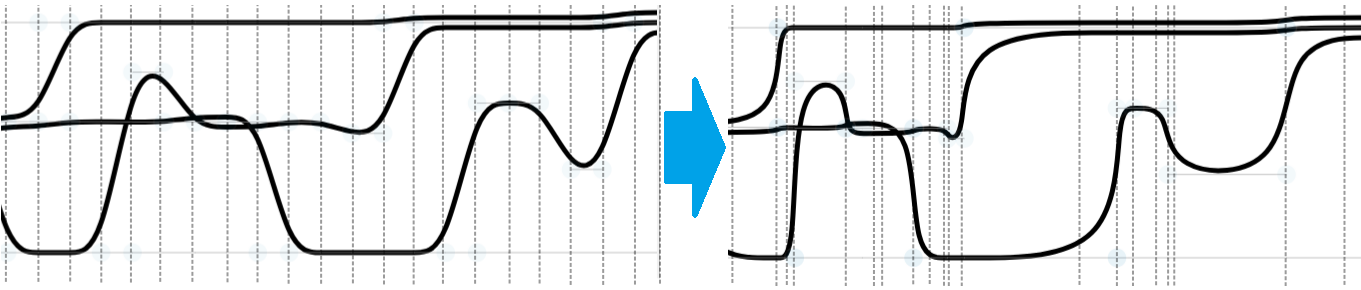}}}
        \caption{Finalization by aligning X position to the exact time.}
        \label{fig:temporal}
          \end{subfigure}
    \caption{Stages of the layout algorithm which transform the event-based data into a final storyline layout used by the interactive system.}
    \label{fig:layout}
\end{figure}

We propose a modified storyline~\cite{SoftwareEvolutionStorylines} layout algorithm to support our tasks. While typical approaches to storyline drawing optimize the number of edge crossings and minimize bends, we have the constraint of patient locations, including ward (Task 3), and patient contacts (Task~2). Fast layout is required for interactive exploration, as sets of patients in the view can change when filters are applied, and data sets are loaded.  Thus, we prioritize runtime over crossing optimization and minimizing bends (see~ \autoref{sc:discussion}).  We build upon existing layouts and combine them and adapt them for our purposes. Given these desired goals, our algorithm is structured as follows (see also \autoref{fig:layout}):

\begin{enumerate}
    \item \textit{Data pre-processing}: Construct patient contact graph
		\item \textit{Initialization}: Compute initial layout
   \item \textit{Constrained force-directed layout}: 
    \begin{enumerate}
        \item \textit{Constraint -- x position}: The temporal order of patient transfers is preserved 
        \item \textit{Constraint -- home/hospital}: Patient y position for home, hospital and temporary \ch{home} location is preserved
		\item \textit {Force -- ward separation}: Different wards are spaced out on the y-axis 
        \item \textit{Force -- ward position}: Spring forces encourage keeping wards at the same y position 
        \item \textit{Force -- line straightening}: Force encourages straight lines after each iteration 
    \end{enumerate}
  \item \textit{Finalization}:
		\begin{enumerate} 
		\item \textit{Line order}: For simultaneous contacts of many patients, y~order of lines according to the transfer time
	\item \textit{Temporal adjustment}: Align x position according to the exact time moments
	    \end{enumerate}
\end{enumerate}

\textit{1. Pre-processing} We construct a directed acyclic graph (DAG) from the event-based data. Nodes are sets of patients at a location (i.e., patient contacts) or infection state changes (see \autoref{fig:preprocess}). Edges are transitions between these states.  As the transfers have a single timestamp, nodes are interpreted as instantaneous events.  Edges that connect these nodes form a from-to transfer edge. This duplication also ensures the duration of the stay at a particular location, and all possible contacts between two transfer time moments. The locations, location changes, and patient contacts can be followed through paths in this graph. For changes in infection state, an infection state change node is created and passing through this node changes the patient's infection state.

\textit{2. Initialization}  As each node of the DAG represents a set of patients together in a location, an initial layout can be computed using dynamic set visualization based on Sankey Diagrams (D3 implementation used~\cite{D3}). This initial layout takes into account the cardinalities of the sets and operates on the structure of the DAG only. The exact time of patient \ch{transfers} is taken into account in the finalization.


\textit{3. Constrained force-directed layout} We employ a force-directed layout for storylines~\cite{SoftwareEvolutionStorylines,silvia2015storyline}. Our initialization helps avoid local minima, but it requires refinement subject to additional constraints (e.g., to convey patient location). 
After experimenting with various ways to use space to convey patient location (i.e., all on the bottom), we decided to divide the screen into home -- hospital -- temporary \ch{home} from top to bottom (see~\autoref{fig:areas}). Admission to the hospital is a vertical line descending from the top area of the screen. Temporary home transfers are vertical lines towards the bottom of the screen.  
When optimizing the layout with the force-directed algorithm, we use a constraint-based approach~\cite{dwyer2006ipsep} to enforce these locations and use additional forces to encourage the desired properties of the layout. The first constraint preserves the temporal order of movements along the x-axis. The second constraint restricts movement outside y-axis areas for the three location types (Task 3.1). Additional forces are used to maximize the stability of the y position of a patient during a stay in a ward to help represent contact location (see~\autoref{fig:layoutwards}) (Task 3.2.).  Still, the y position can vary significantly when a patient stays in a location for long periods of time. To remove this position change, we replace the individual y positions of these nodes with the average (see~\autoref{fig:straightening}). 

\textit{4. Finalization} \ch{L}ines are ordered from top to bottom according to the order patients entered the ward (see~\autoref{fig:ordering}). Even though this may cause more edge crossings, this order helps support Task 2\ch{: tracing transmissions. P}atients that were in the ward for longer periods of time are more likely to be involved in transmission events. \ch{Finally}, individual nodes are placed at the precise time of their events along the x-axis, which is important for determining outbreak duration and time of possible pathogen transmissions (Task 4) (see~\autoref{fig:temporal}).

\subsection{Visual Design and Interaction}
\label{sc:storylinedesign}

\begin{figure}[tbhp]
    \centering
    \includegraphics[trim=0 0 50 00, clip,width=1\linewidth]{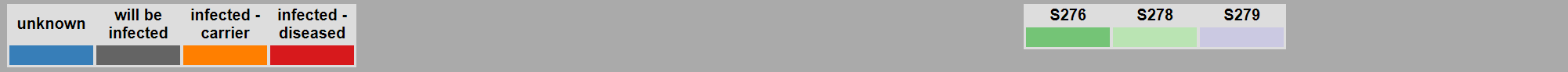}\\
        
             \textcolor{gray}{
 \setlength{\fboxsep}{0pt}%
\setlength{\fboxrule}{0.5pt}%
\fbox{\includegraphics[trim=0 0 0 00, clip,width=0.45\linewidth,height=3.5cm]{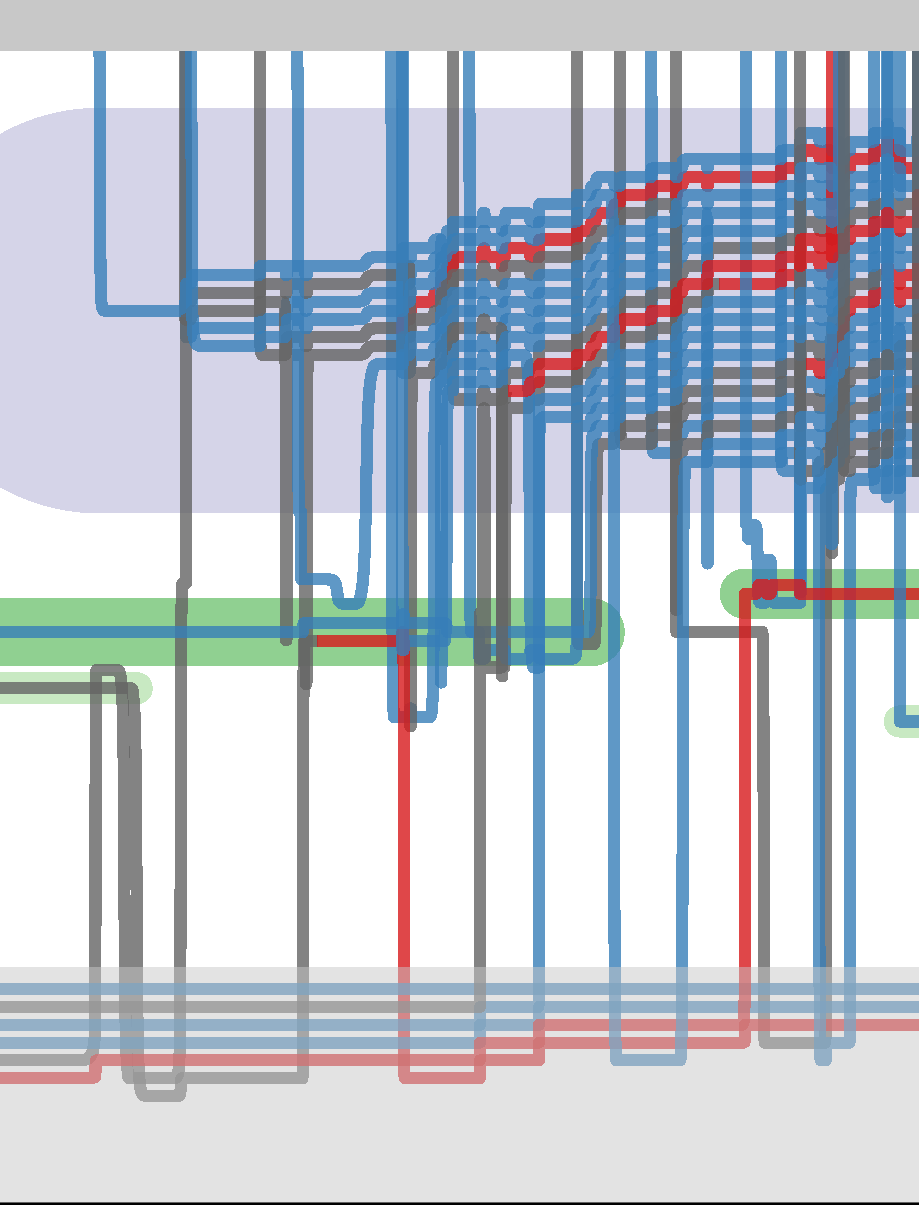}}}
 \hspace{1mm}
        \includegraphics[width=0.06\linewidth]{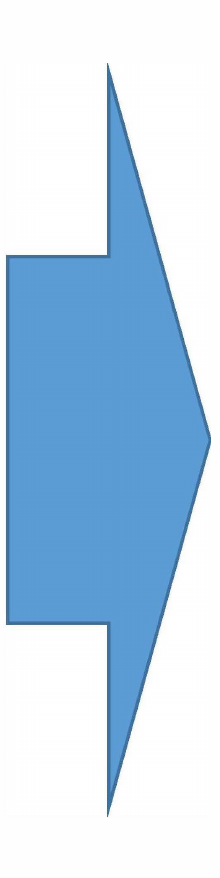}  
\textcolor{gray}{
 \setlength{\fboxsep}{0pt}%
\setlength{\fboxrule}{0.5pt}%
\fbox{\includegraphics[trim=0 0 0 10, clip,width=0.45\linewidth,height=3.5cm]{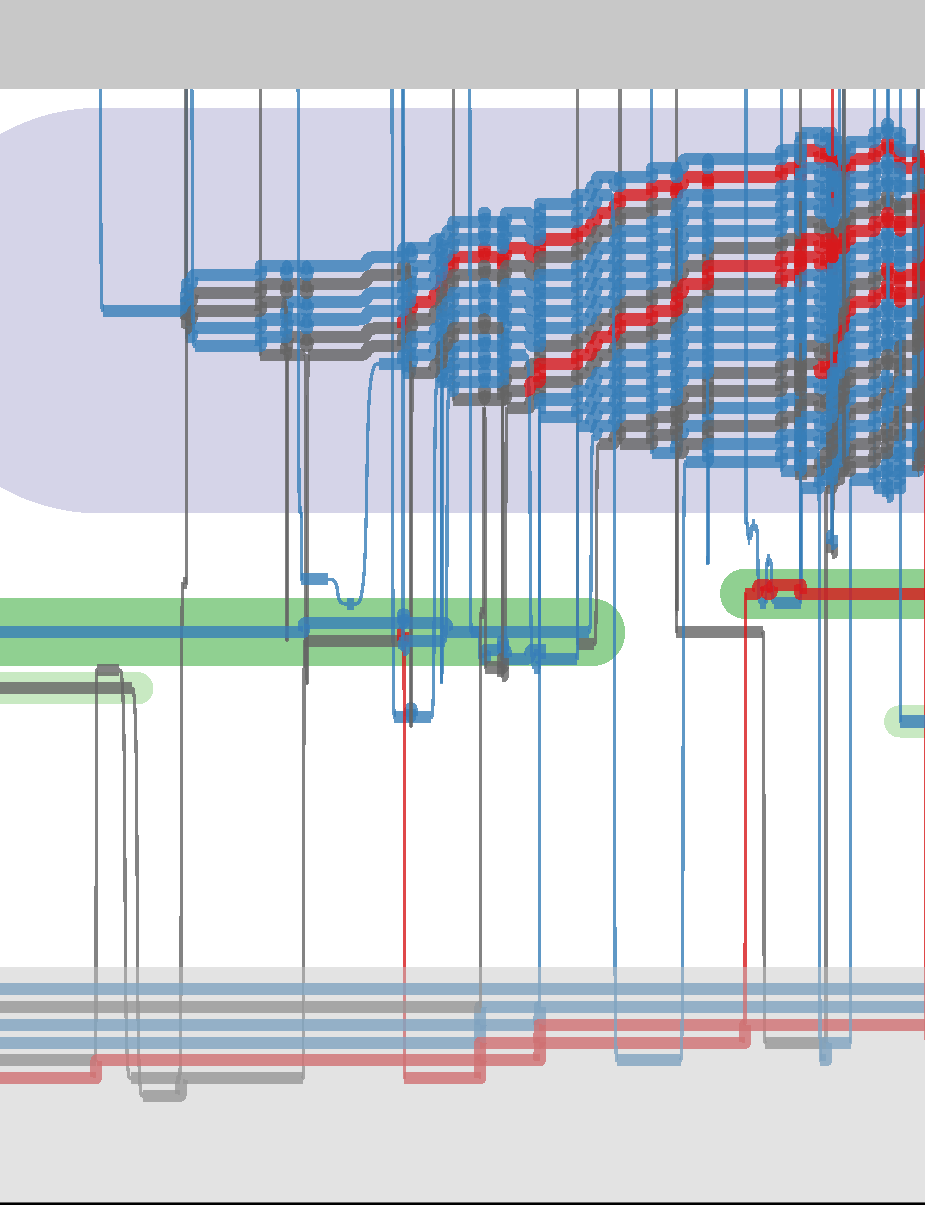}}}
   \caption{\ch{Left: Visual design, where line color encodes infection status. Colored background hulls display wards. Right: Emphasizing patient contacts by reducing the width of vertical lines}.}
    \label{fig:visualdesign}
\end{figure}

Infection status is encoded using color (see \autoref{fig:visualdesign} left) in order to help in tracing transmissions (Task 2). Differentiating between `unknown' and `unknown-will be infected'  helps infection control experts track patients over long time periods, reducing the requirement to pan and zoom. Details on microbiological data are shown on demand through a tooltip. The contact location (Task 3), specifically the ward, is shown on demand by a colored background hull. 

Using a process inspired by~\cite{tang2018istoryline,kim2010tracing}, patient lines are drawn smoothly. As short periods of contact can lead to transmission, wiggles in the line indicate new contacts (see \autoref{fig:ordering}).  \ch{We emphasize contacts between patients on the same ward (Task 2) by minimizing the line width for vertical lines (see \autoref{fig:visualdesign} right). This also reduces overplotting as inspired by Tanahashi and Ma~\cite{tanahashi2012design}.}

Standard pan and zoom operations are supported as well as highlighting of selected patients. When a patient is selected, its line width is increased. Non-focus information is not filtered out~\cite{kim2010tracing}, as it is a requirement to be able to track patients in the data set. Straightening selected patient lines~\cite{liu2013storyflow} would be an alternative, but this would interfere with our ward location encoding.

\subsection{Support for Transmission Pathway Exploration}
\label{sc:pathwayexploration}

Potential pathogen transmissions must be visible~\cite{munznerTJ} among many patient contacts over long periods of time (months to years). Panning and zooming over such long time periods is inefficient and may lead to missing important contacts in the data. We developed specialized tracing interactions to support this visual analysis.

Our interactive approach supports 1) \textit{backward tracing} -- finding transmission events and patients in the past that could have infected a selected patient. The interaction \ch{enables} the search for an index patient, i.e., patient zero (Task 2). 2) \textit{forward tracing} -- finding patients that could be infected by a selected infected patient at a \ch{later point in time}, i.e.,  putative infected patients (Task 5). 

\begin{figure}[t]
    \centering
    \begin{subfigure}{\linewidth}
       \includegraphics[width=\linewidth]{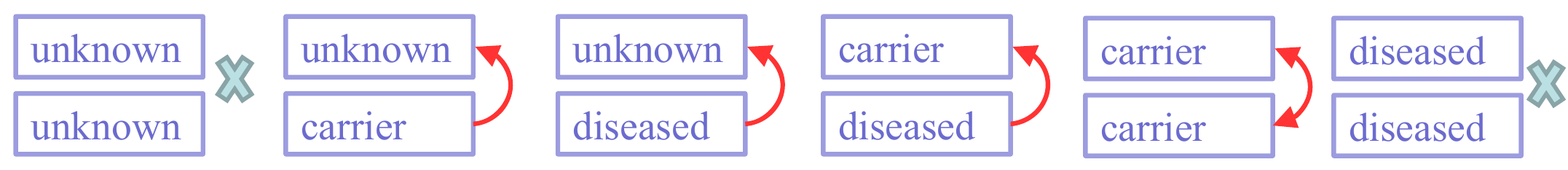}
        \caption{Contact event types for infection transmission}
    \end{subfigure}
    \begin{subfigure}{\linewidth}
        \includegraphics[width=\linewidth]{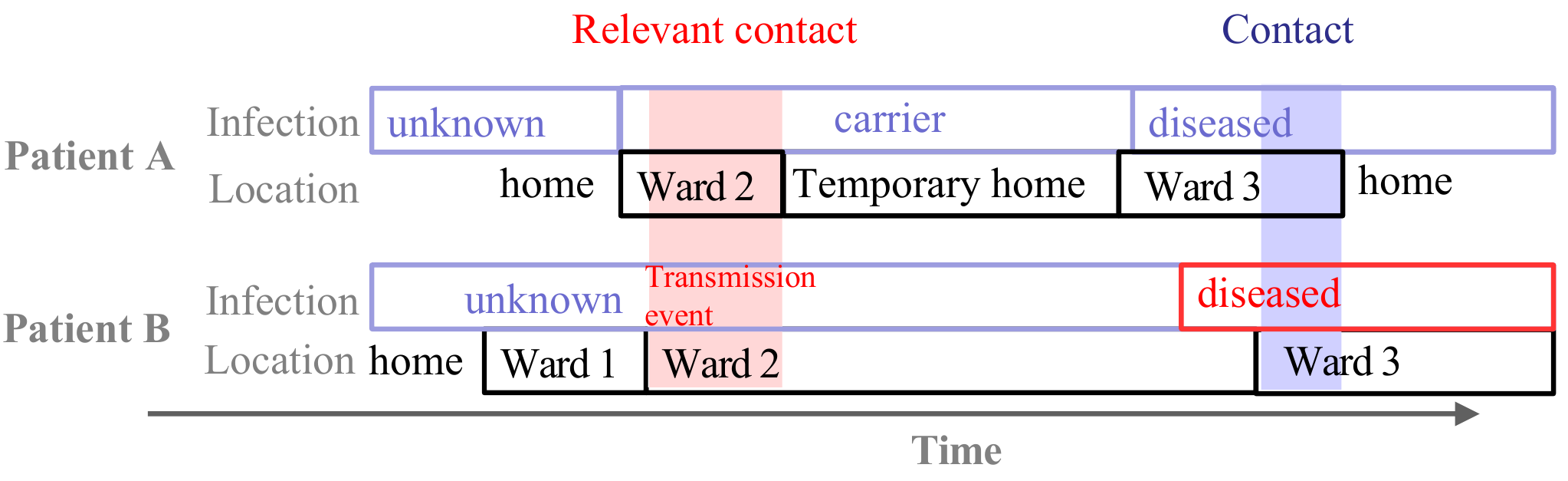}
        \caption{Relevant contacts for pathogen transmission and possible transmission events.}
    \end{subfigure}
    \caption{The types of contacts and encoding of transmission events.}
    \label{fig:infectionstate_transmission}
\end{figure}

We now explain the necessary computation required to interactively backward trace. Computation for forward tracing is done analogously. For a selected patient $P$, we calculate which contacts might have lead to the patient's infection. A pathogen can be transmitted from an infected patient $P_i$ to the patient $P$ when the patients come into contact. \ch{Note, a} contact between two diseased patients is deemed irrelevant as both patients are already diseased (see \autoref{fig:infectionstate_transmission} top). We identify relevant contacts as shown in \autoref{fig:infectionstate_transmission}: the pathogen transmission must have happened during a \textit {critical contact} before the infection was detected -- the time moment of the earliest positive microbiological result $\tau$. 
For each $P_i$ of $P$ and each contact location $L$, we determine the earliest critical contact before the infection is detected $CC_{i,L}=min_{t_s}\{P_j, t_s, t_e, t_s\leq \tau:L(P)=L(P_i)=L)\}$. We assume the starting time moment of the contact interval $(t_s,t_e)$ where $t_t=t_s$ as the transmission event time, $L_t$ is transmission location (Task 3) and $P_t$ is transmission contact. Note, $P$ can have several potential transmission events -- different persons and locations and different times. This analysis is repeated, especially for contacts with \textit{unknown} infection status, as they could be potentially infected if they had  contact to an infected patient before. These critical contacts are computed using a constrained path search in the DAG used for the layout. There is currently no bound on how far back or forward in time the transmission events are searched, \ch{but user-specified bounds could be implemented, depending on the specifics of the investigated pathogen}.

\begin{figure}[tbp]
\centering
\includegraphics[trim = 0 0 0 0,clip,width=\linewidth]{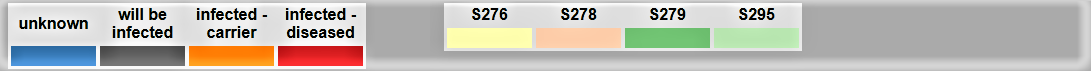}\\
\textcolor{gray}{
 \setlength{\fboxsep}{0pt}%
\setlength{\fboxrule}{0.5pt}%
\fbox{ 
\includegraphics[trim=250 0 0 50, clip,width=\linewidth]{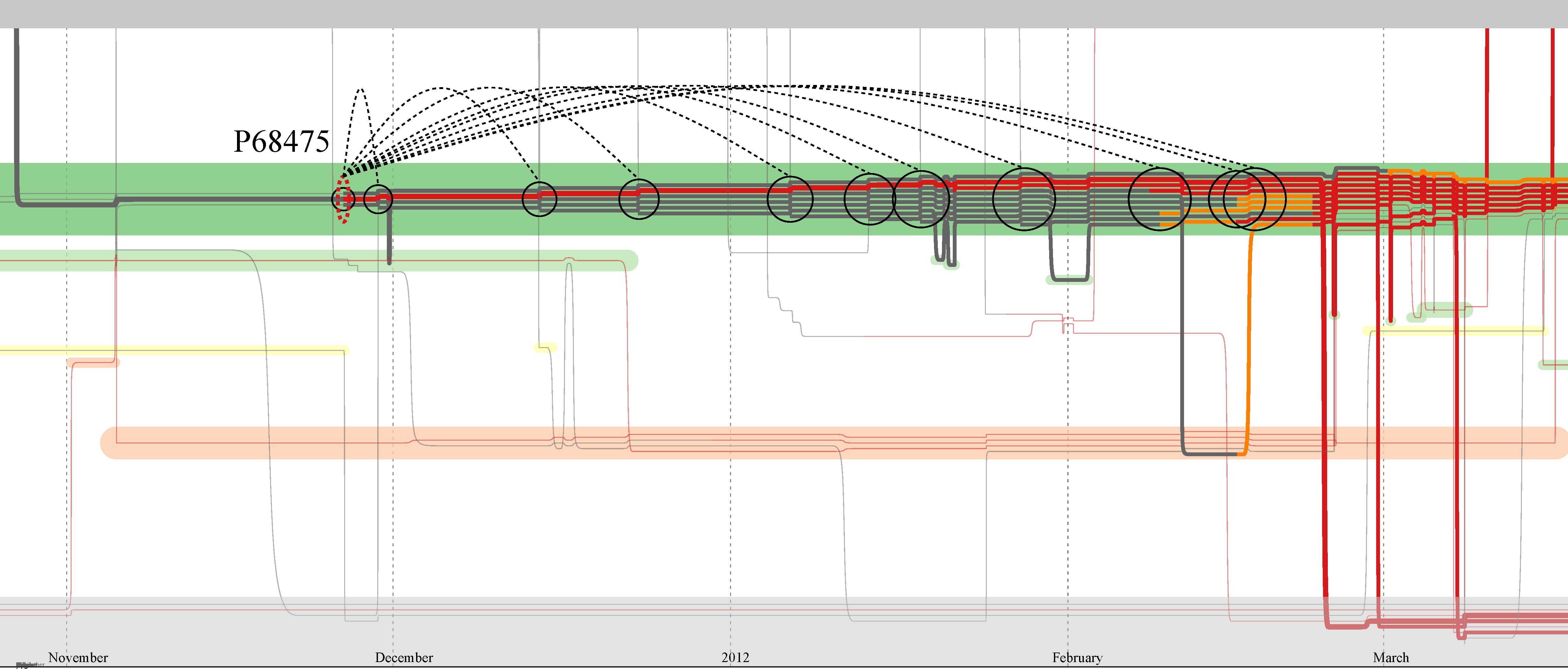}}}
 \caption{Forward tracing of pathogen transmissions of patient P68475. The patient potentially infected many patients at the ward S279.}
 \label{fig:forwardtracing}
\end{figure}

\begin{figure*}[tbp]
\centering
\includegraphics[trim = 0 0 300 0,clip,width=\linewidth]{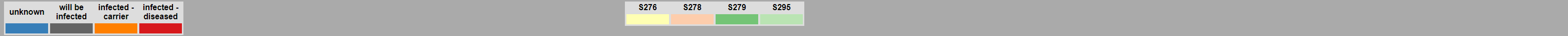}\\
\textcolor{gray}{
 \setlength{\fboxsep}{0pt}%
\setlength{\fboxrule}{0.5pt}%
\fbox{ 
\includegraphics[trim=0 20 0 50, clip,width=\linewidth,height=5.5cm]{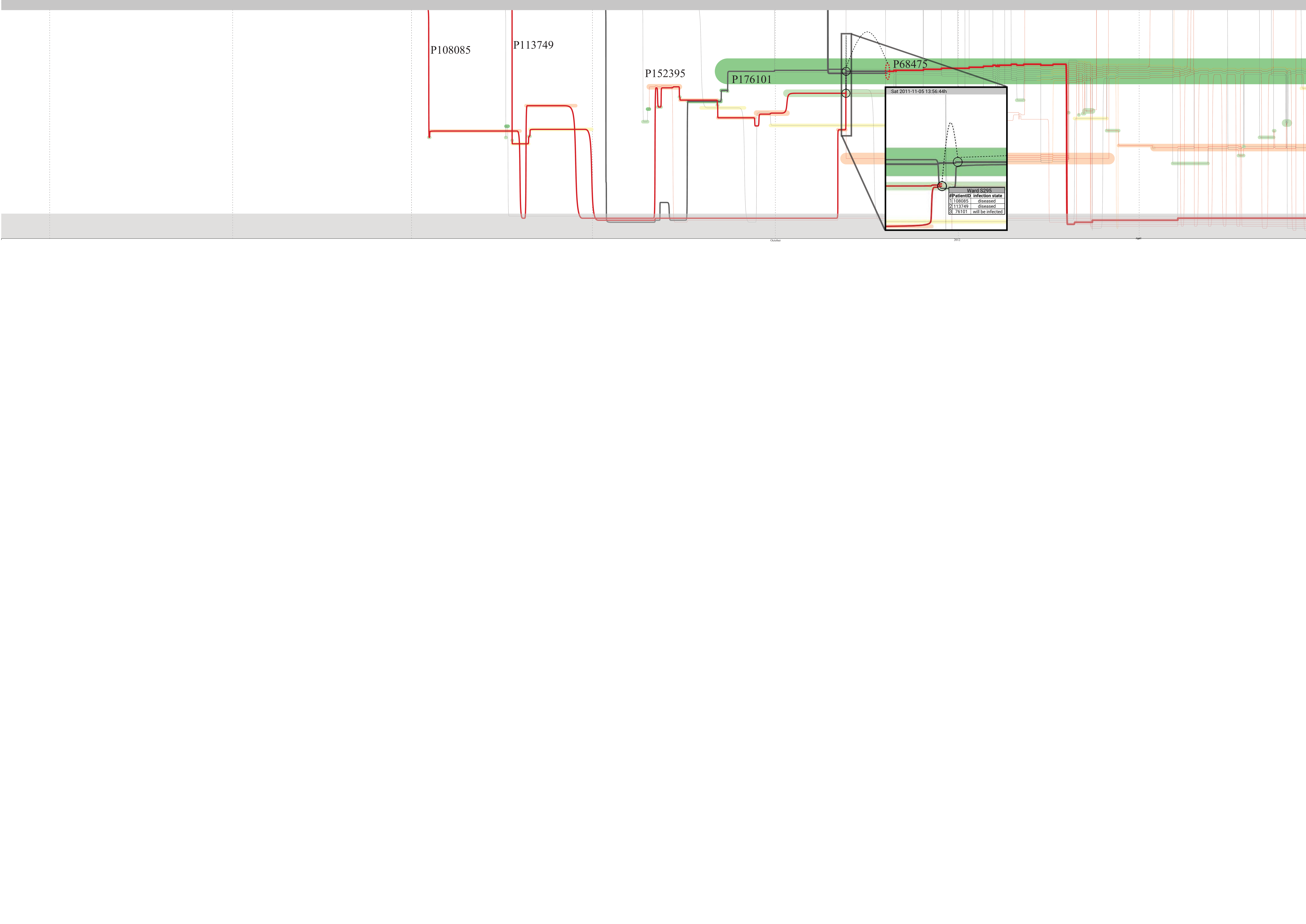}}}
        \vspace{-2mm}
 \caption{Backward tracing of pathogen transmissions of patient P68475 -- indirectly via patient P76101's contact with two patients on ward S295.}
 \label{fig:indirecttransmission}
\end{figure*}

In the view, the selected patient $P$ is emphasized. Relevant contact patients $P_t$ are kept and the non-relevant patients are de-emphasized. The positive test/screening events $\tau$ are highlighted with ellipses (orange for carrier and red for diseased) and connected to potential transmission events $(t_t,P_t,L_t)$ with flashback lines inspired by~\cite{padia2019system} (see~\autoref{fig:forwardtracing} \&~\ref{fig:indirecttransmission}). The circle color encodes the type of relevant contact (Task 2.1). We show all possible transmission events -- i.e., several flashback lines. Connections indicate the length of time the potential infection was not detected by the screening or testing (Task 4). 


\section{Case Study}
\label{sc:usecase}

The visual analytics system was applied to real data from a large German hospital by an infection control expert. The anonymized  data includes hospital location data of  $\sim$180,000 patients and $\sim$900,000 microbiological data over four years.
The focus of the case study is a multi-resistant pathogen of special interest: the bacteria \klpn. \klpnShort is a commensal gut bacterium that is a dominant cause of hospital-acquired infections. It is responsible for infections in the urinary tract, respiratory tract and blood stream~\cite{magill2014hai}. 

In the first quarter of 2012 (actual date anonymized), the infection control experts faced an outbreak of \klpnShort in an \ch{area with four wards}. Staff members work on all four \ch{wards} and the patients are frequently swapped between these locations. The total capacity of the \ch{area} is 50 beds. The outbreak investigation performed at that time was done manually by the infection control experts, meaning the data was collected from several systems and merged into spreadsheets. The original analysis using classical epidemiological data assigned in total 12 patients to the outbreak. However, a deeper analysis of the pathogen transmission was not possible. 

\ch{This use case shows how the infection control experts analyzed this outbreak retrospectively using the system. The interaction with the system was supported by the visual analytics experts. Since the \textit{patient zero} was not found during the first outbreak investigations, the experts aimed to trace back the initial source of infection \ch{and to} identify potential cases that were overlooked at that time.}

\begin{figure}[btp]
    \centering
   \begin{subfigure}{\linewidth}
       \textcolor{gray}{
 \setlength{\fboxsep}{0pt}%
\setlength{\fboxrule}{0.5pt}%
\fbox{\includegraphics[trim=5mm 0 0 0, clip,width=\linewidth]{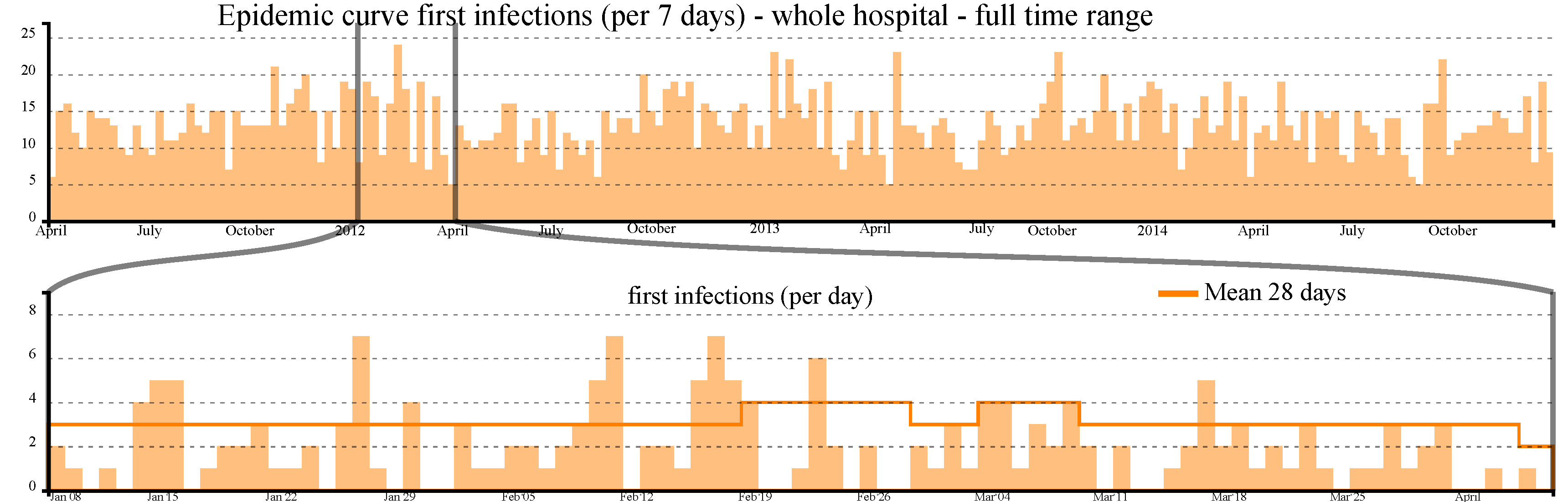}}}
         \caption{Epidemic curve for new infections of \klpnShort -- sum of whole hospital.}
         \label{fig:UCEChospital}
    \end{subfigure}
       \begin{subfigure}{\linewidth}
        \textcolor{gray}{
 \setlength{\fboxsep}{0pt}%
\setlength{\fboxrule}{0.5pt}%
\fbox{\includegraphics[width=0.48\linewidth]{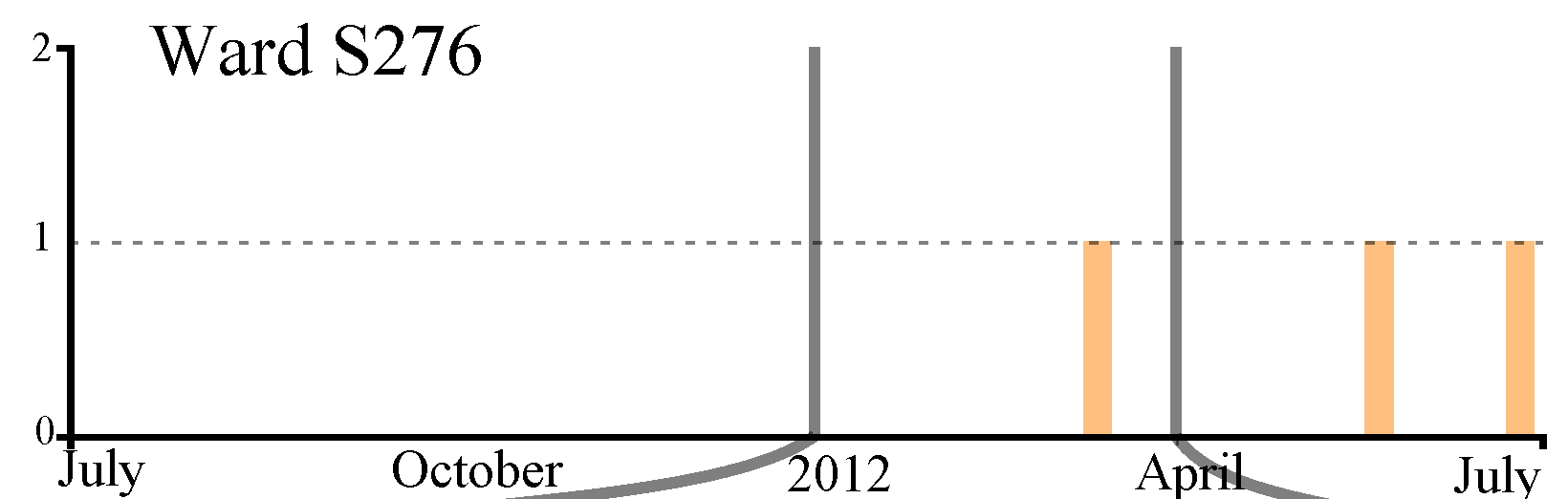}}}
        \textcolor{gray}{
 \setlength{\fboxsep}{0pt}%
\setlength{\fboxrule}{0.5pt}%
\fbox{\includegraphics[width=0.48\linewidth]{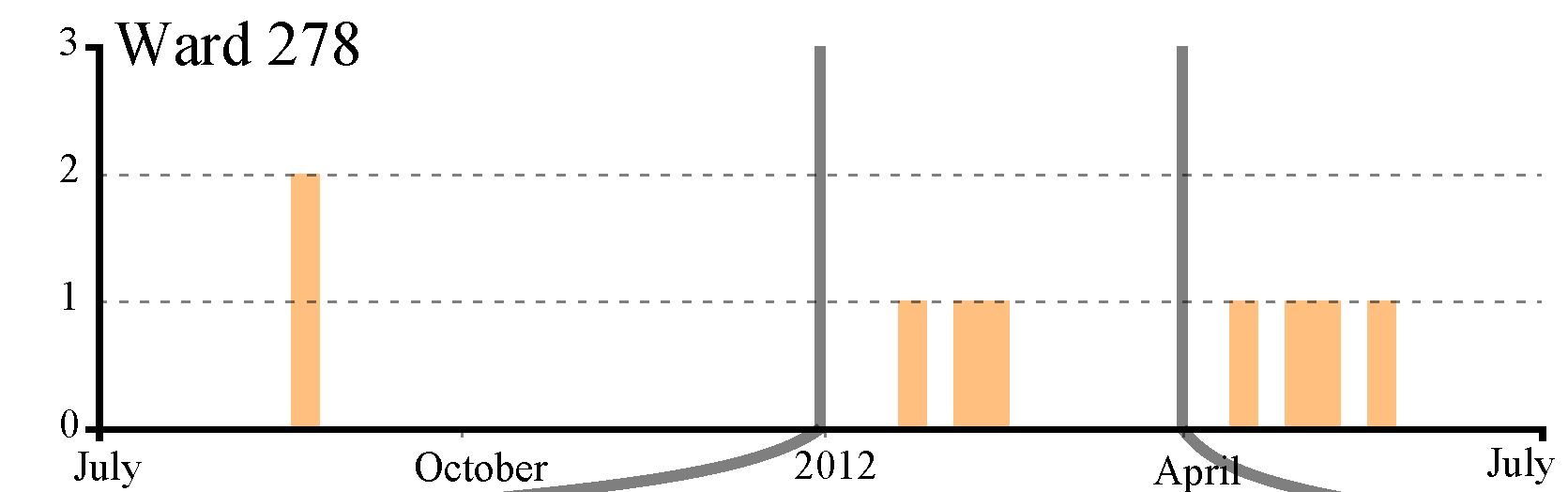}}}\\ 
       \textcolor{gray}{
 \setlength{\fboxsep}{0pt}%
\setlength{\fboxrule}{0.5pt}%
\fbox{ \includegraphics[width=0.48\linewidth]{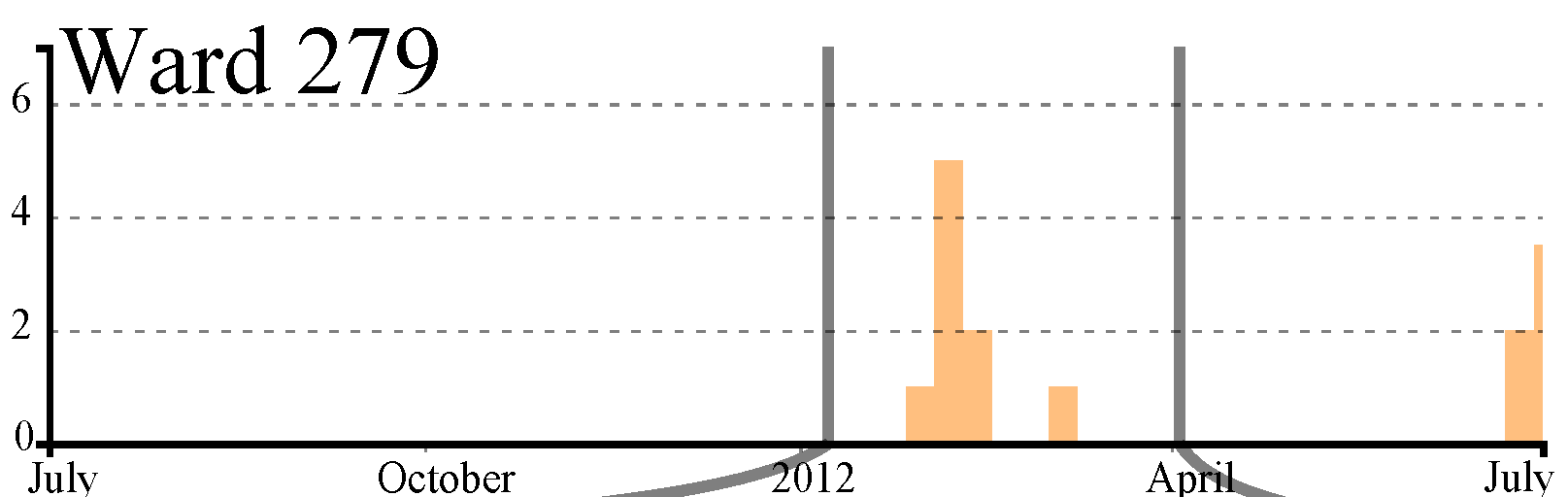}}}
       \textcolor{gray}{
 \setlength{\fboxsep}{0pt}%
\setlength{\fboxrule}{0.5pt}%
\fbox{ \includegraphics[width=0.48\linewidth]{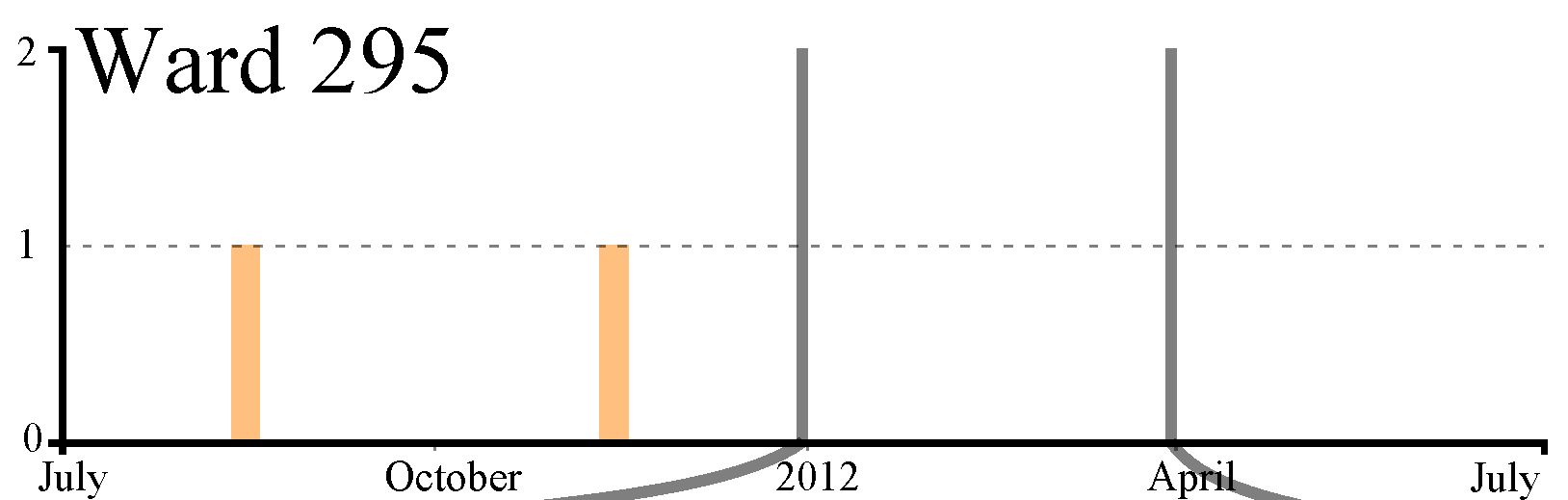}}}
        \vspace{-2mm}
         \caption{Epidemic curve for new infections of \klpnShort -- suspicious wards.}
         \label{fig:UCECwards}
    \end{subfigure}
    \caption{Epidemic curve with focus on the time period of interest: Early-2012 indicates an higher number of new infections at ward S279.}
    \label{fig:UCEC}
\end{figure}

First, the infection control expert consults the epidemic curve for \klpnShort of the first months of 2012 for the entire hospital to assess the number of infected patients (Task~1). The data was cleaned for copystrains (repeated samples of the same patients), allowing to focus on new positive results (see \autoref{fig:UCEChospital}). Only a small increase in the number of infections at the hospital has occurred. Thus, this outbreak is difficult to detect automatically. Based on his expertise, his focus turns to the four wards, S276, S278, S279, and S295 (see \autoref{fig:UCECwards}), where the outbreak was initially localized. The numbers of new infections in February in the wards were: 1 patient (S276), 3 patients (S278) and 9 patients (S279) (see \autoref{fig:UCECwards}). The number of newly infected patients on the wards was higher than the endemic level assuming an \klpnShort outbreak on the ward.

The expert begins to reconstruct the transmission pathway (Task~2) and visualizes all positively tested patients on the wards in the \storyline (see supplementary material and~\autoref{fig:forwardtracing}). In total, there were 23 infected patients, which is higher than the original outbreak analysis showed. A dominant cluster of patients highlights 12 infected patients on ward S279 in winter 2011/2012. It reveals the first infected patient, patient P68475, who was already tested positive in November 2011 -- three months before this cluster of \klpnShort (Task~4).  
By using the \textit{forward tracing interaction} from patient P68475, the infection control expert identified further patients that were in contact with this patient (see \autoref{fig:forwardtracing}). Thus, patient P68475 is the cause for the spreading of \klpnShort on ward S279, but not for all patients in the outbreak.

\begin{figure}[tb]
\centering
\includegraphics[trim=0 0 500 0,clip,width=\linewidth]{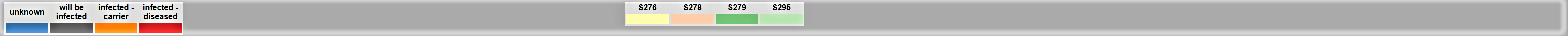}\\
\textcolor{gray}{
 \setlength{\fboxsep}{0pt}%
\setlength{\fboxrule}{0.5pt}%
\fbox{ 
\includegraphics[trim=300 0 100 50,clip,width=\linewidth]{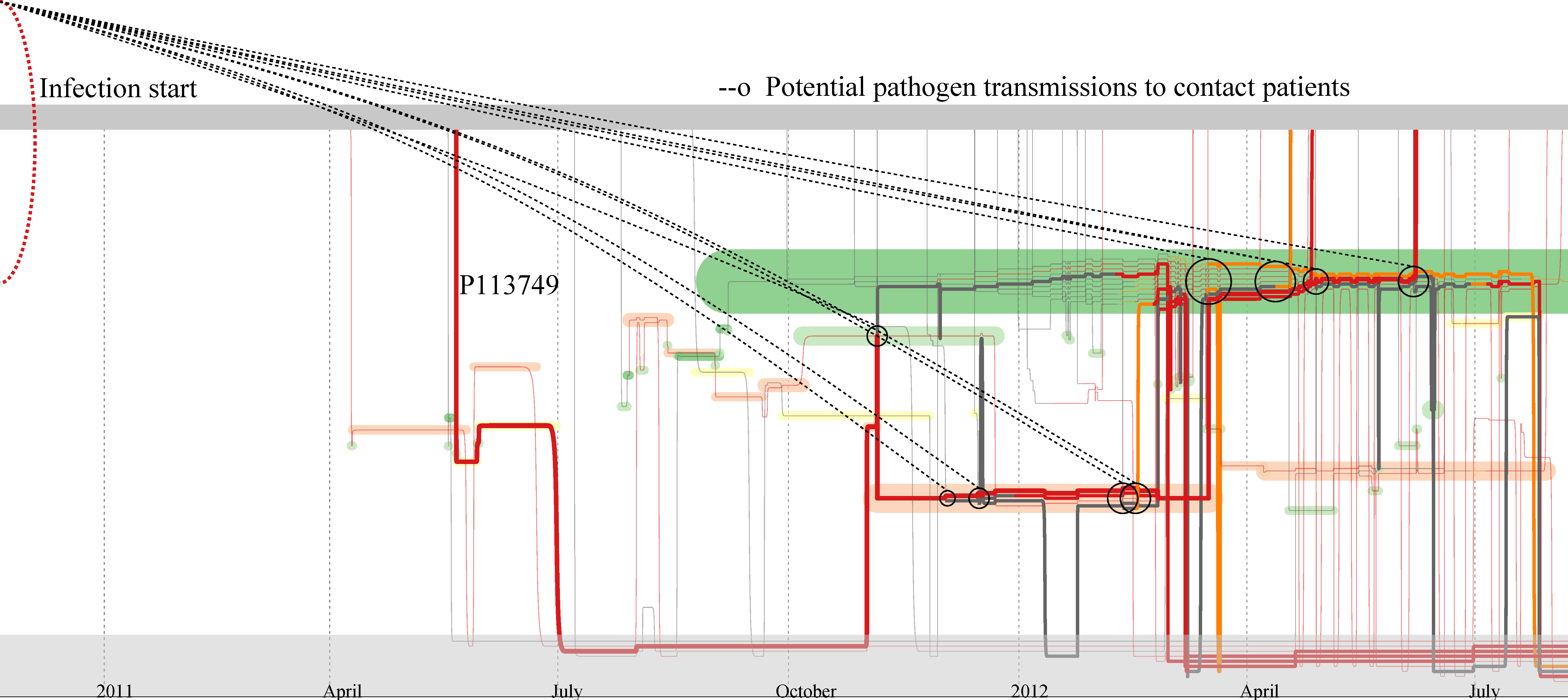}}}
    \caption{Forward tracing for  P113749 shows potential early start of outbreak and transmissions to patients at wards S279, S295 and S278.}
    \label{fig:indirecttransmissionFF2pat}
\end{figure}

\begin{figure}[tb]
\centering
\textcolor{gray}{
 \setlength{\fboxsep}{0pt}%
\setlength{\fboxrule}{0.5pt}%
\fbox{ 
\includegraphics[trim=50 0 500 130,clip,width=0.8\linewidth]{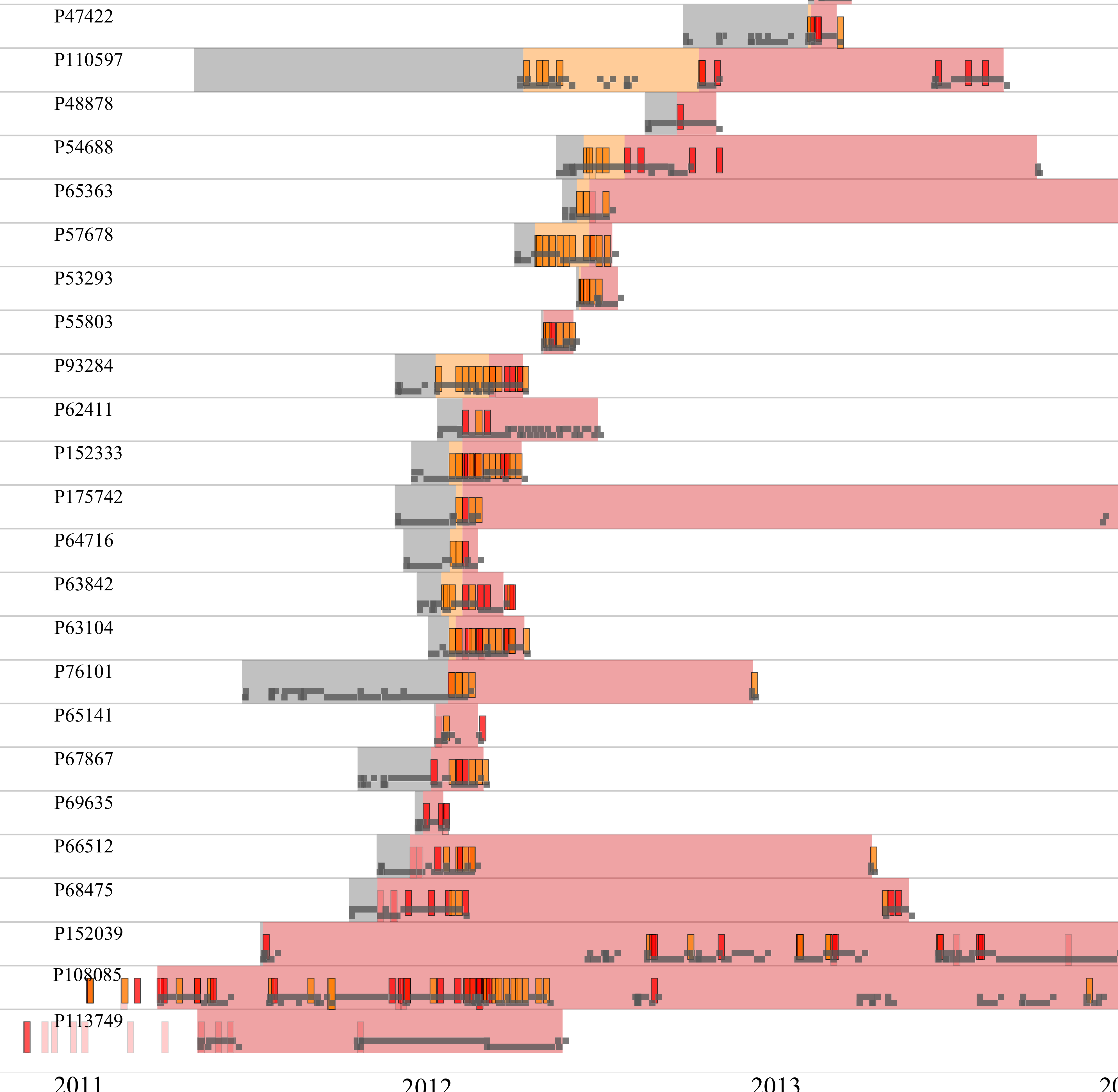}}}
    \caption{\linelist of the outbreak patients sorted by time of first infection shows first infected patients already in 2011. The red background indicates the infection status starting from the first laboratory evidence of \klpnShort --shown as red vertical bar.} 
    \label{fig:UClinelist}
\end{figure}

Since there was no previous contact of patient P68475 with colonized patients (carrier) or diseased patients (with \klpnShort) in ward S279, the expert uses the \textit{backward tracing interaction} to search for potential contacts on other wards (see \autoref{fig:indirecttransmission}). The visualization showed that patient P68475 had no direct contact to patients with positive laboratory tests for \klpnShort. However, the tracing lines lead to ward S95. As shown in the zoomed view of \autoref{fig:indirecttransmission}, patient P68475 had contact with infected patients indirectly via patient P76101, who, in turn, was in contact with two patients on ward S295 (Task~2.1). This is a novel insight. Both patients, P113749 and P108085, were tested positive for \klpnShort. Thus, a possible transmission route of \klpnShort to ward S279 could be via $P113749/108985 - P761010 - P68475$. As \autoref{fig:indirecttransmission} shows, both patients are frequently returning to the hospital (many vertical lines from bottom) and usually stay for longer periods in the hospital (long horizontal lines). Forward tracing shows the potential beginning of the transmission events was before their first admission to the hospital (see the start and the length of the tracing lines in \autoref{fig:indirecttransmissionFF2pat}). \linelist confirms this finding (see \autoref{fig:UClinelist}), where the first positive microbiological results (red vertical lines of patient P113749 and P108985) are before the first hospital admission (Task~3.1). This is due to the limited dataset availability. The real first admission was before the start of the available data. The forward tracing shows that both patients could infect patients at ward S295 and S279 (see \autoref{fig:indirecttransmissionFF2pat}).  

The \linelist (see \autoref{fig:UClinelist}) identifies a third infected patient, patient P152039 on ward S276, who tested positive before the increase of infected patients occurs (see red vertical line in 2011). Since all patients had contact to several patients, these three patients are regarded as \textit{patient zero(s)} and the potential origin for the \klpnShort for this outbreak on the four wards S276, S278, S279 and S295 (Task~2\&~3) with genetic analysis required for confirmation.

\ch{To provide further support and verification for the hypothesized transmission pathways, the experts also used genome information of the pathogen. Comparison of genomes of bacterial strains adds further evidence to the analysis of outbreaks. Details of this external analysis are provided in the annex in the supplementary material. The genome analysis confirms that eleven patients on the wards S279 and S295 were infected by one \klpnShort bacteria strain, including the potential \textit{patient zero} candidate P113749. Interestingly, the two other potential \textit{patients zero} P108085 and P152039 reveal substantial genome differences to this set of patients (see Fig.~A1 in annex). Thus, the expert assigns P113749 as \textit{patient zero} of this outbreak on the wards. The genome analysis reveals P1520395 as a further source of infection for at least one more patient on the wards. Thus, the two transmission pathways occurred simultaneously (Task~2.3). This is a novel insight.}

\begin{figure}[tb]
\centering
\textcolor{gray}{
 \setlength{\fboxsep}{0pt}%
\setlength{\fboxrule}{0.5pt}%
\fbox{ 
\includegraphics[trim=20 0 0 0,clip,width=1\linewidth]{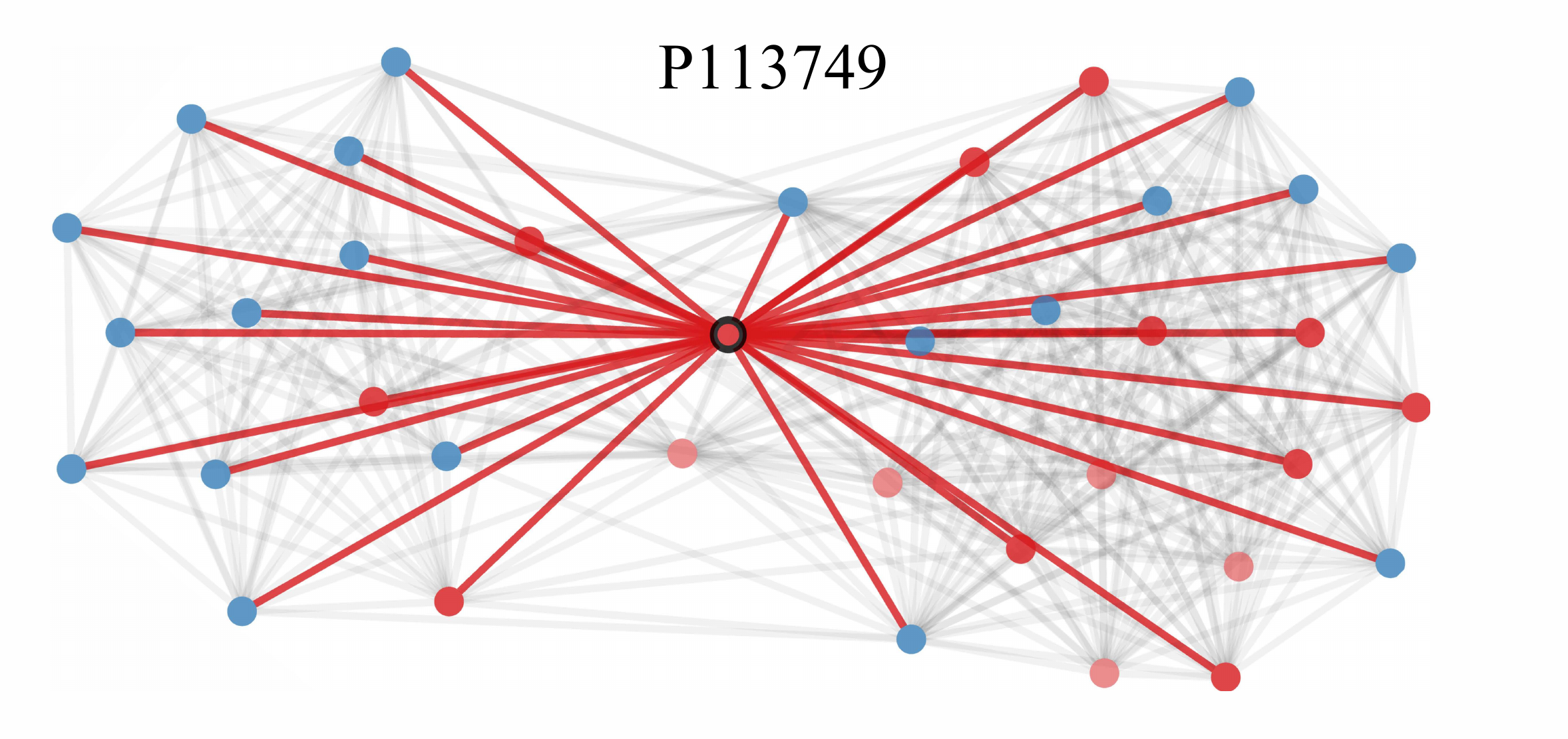}}}
    \caption{ \contactview:  Patient 113749 could have infected more patients in early November 2011.} 
    \label{fig:UCkontaktnetzwerk}

\end{figure}

A further question is how many undetected transmissions the \textit{patient zero} P11349 had and whether there are potentially other infected patients (Task~5). Therefore, the infection control expert opens a \contactview for this patient in the period when the initial transmissions were detected -- beginning of November 2011 (see \autoref{fig:UCkontaktnetzwerk}). The tracing interaction in this view highlights several potential transmissions to patients that were (not) detected as infected during their stay on the wards. \storyline and \linelist (see \autoref{fig:teaser}) show that the non-infected patients were screened (e.g., rectal screening, urine, catheters) regularly and tested negative. 

The use case provides support that the visual analytics system enables the infection control experts to conduct a classical epidemiological analysis in a fast, reliable and comprehensive way. Highlighting potential transmission events enables  to reveal transmission pathways within the hospital. \ch{The steps taken for this analysis with our tool took the hygienist
30 minutes. He estimated that this analysis would normally
take about two working days using his previous methods. The hygienists
usually determine the epidemic curves and contact networks
manually through raw data from multiple information systems.}

\newcommand{\mode}[1]{($Mo={#1}$)}

\section{Domain Expert Feedback}
\label{sc:feedback}

This qualitative evaluation focused on the \storyline through an online questionnaire.  Twenty-five domain experts from seven institutions in Germany participated in this evaluation.
All participants had several years of practical experience with infection control in hospitals: eleven were clinicians, epidemiologists or hygienists, six were medical data analysis experts, six were medical data experts, and two were healthcare managers. \ch{Overall, 60\% of the participants were familiar with our interactive visualization, i.e., have participated in feedback sessions, or had seen a live demonstration of the tool. At the beginning of the experiment, a one- paragraph explanation of each view was provided with no further training.}

\begin{figure}[tb]
    \centering
    \includegraphics[width=\linewidth, trim = {0 0 0 0}, clip]
        {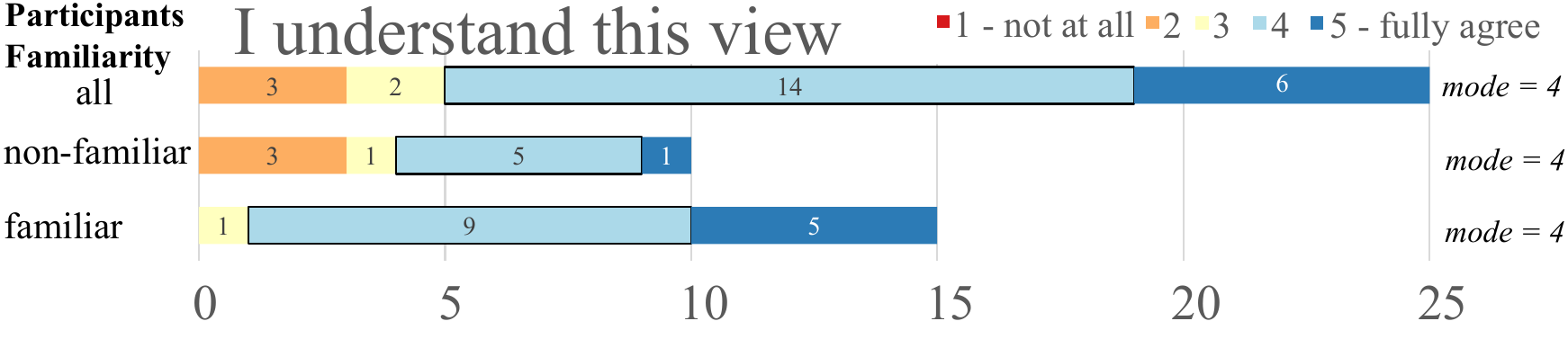}
    \includegraphics[width=\linewidth, trim = {0 0 0 0}, clip]
        {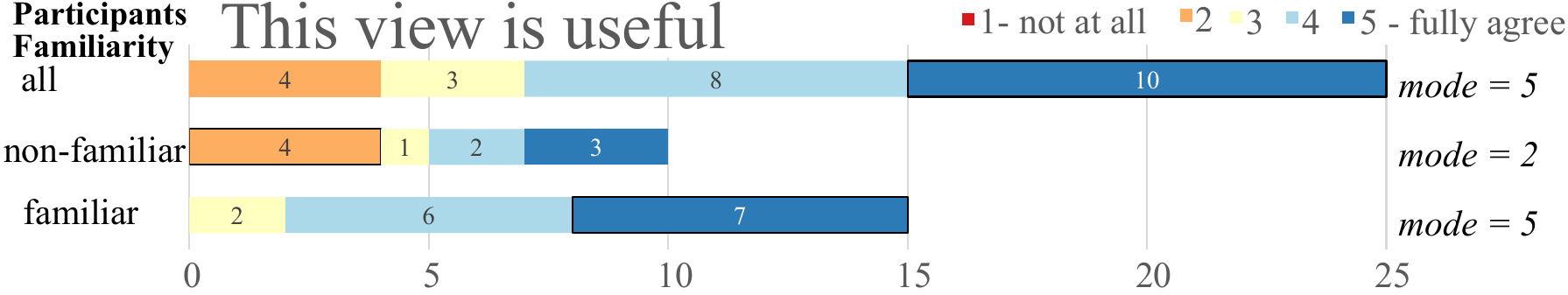}
    \caption{Understandability and usefulness of the view.} 
    \label{fig:eva:base}
\end{figure}

\begin{figure}[tb]
    \centering
    \includegraphics[width=\linewidth, trim = {0 0 0 0}, clip]
        {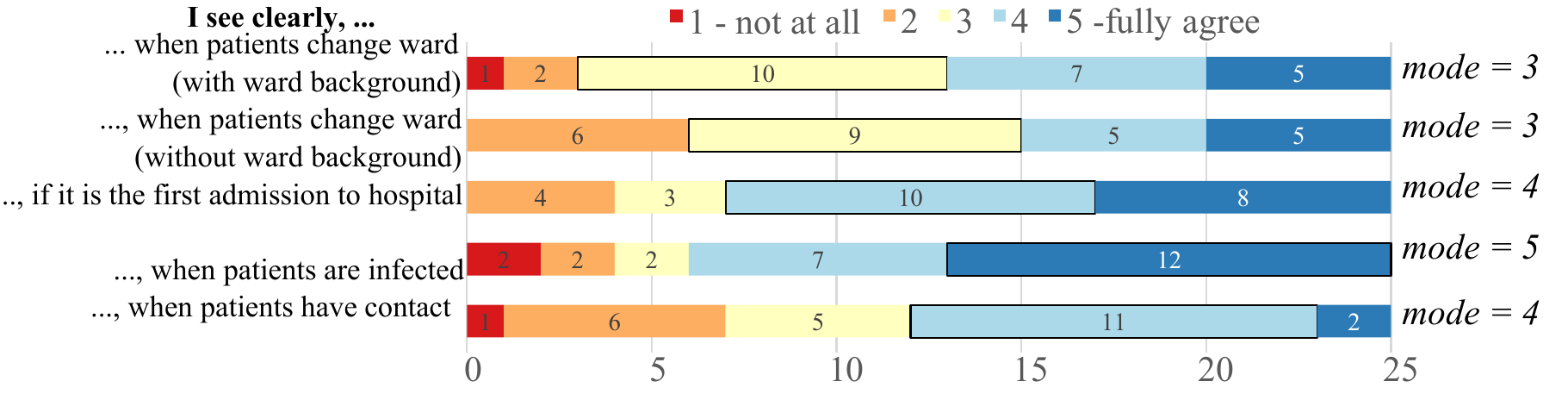}
    \caption{Results for readability of the views.} 
    \label{fig:eva:tasks}
\end{figure}

\begin{figure}[tb]
    \centering
    \includegraphics[width=\linewidth, trim = {0 0 0 0}, clip]
        {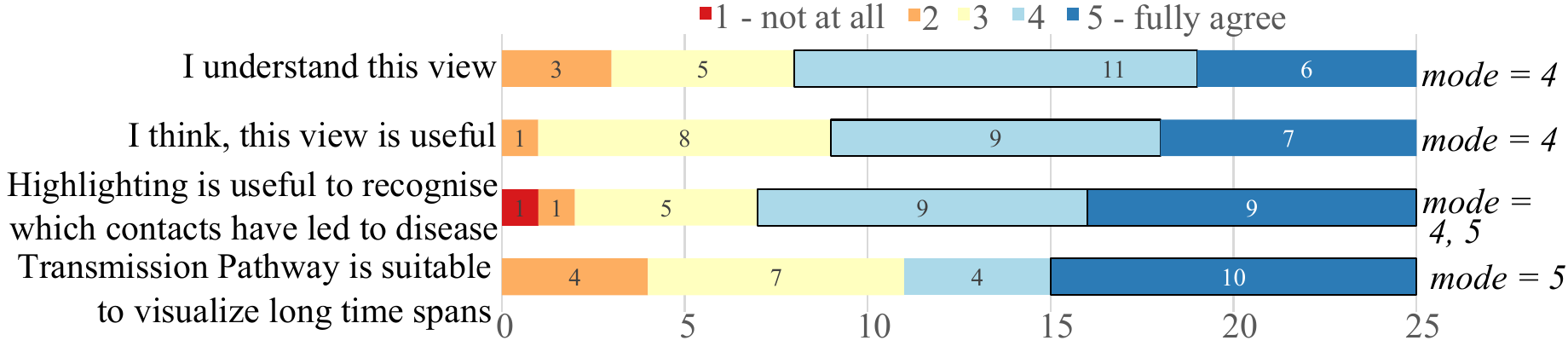}
    \caption{Results for questions on tracing interaction.} 
    \label{fig:eva:highl}
\end{figure}

\ch{The online questionnaire showed the views from the use case and asked participants to read these visualizations and answer the usability and understandability questions as presented in~\autoref{fig:eva:base}--\ref{fig:eva:highl}. Participants rated their answers to each question on a Likert scale: 1--5, with 5 meaning 'fully agree' and could provide free text comments.}

Participants found the view useful \mode{5} and easy to understand \mode{4} (see \autoref{fig:eva:base}).
As expected, participants familiar with the tool gave higher scores.
This result together with free text feedback indicates that training is needed for using the approach effectively.

We also assessed how well the views supported the intended tasks (see \autoref{fig:eva:tasks}).
Participants could identify contacts \mode{4} and understand their infection status \mode{5} (Task 2.1).
They can clearly identify patient stay duration and movement in or out of the hospital \mode{4} (Task 3.1).
Participants found it harder to identify the wards \mode{3} (Task 3.2).  
The free text showed mixed reviews of using background color with some participants preferring it while others not. Therefore, we interactively enable/disable this encoding.


Participants found the highlighting of potential transmission events very useful \mode{4} and understandable \mode{4} (see \autoref{fig:eva:highl}).
They found the feature helpful in determining which contacts could lead to a transmission and spreading of pathogens \mode{4,5}.
The feedback on support for long time periods when determining outbreak duration (Task 3) was also positive \mode{5}.
In free text, experts called for further pan and zoom support, which we revised in our visual interface.

The \textit{comparison of the network diagram and pathway views} showed that nine participants preferred \storyline and six participants preferred \contactview.
Ten participants preferred a combination of the two views. \ch{The distribution of preferences was the same for both familiar and non-familiar participants.}
This result is expected as the views target different tasks:
``I need both views, as they help find an answer to different tasks.'' 
\storyline is useful for detailed spatio-temporal analysis of transmissions and the \contactview is useful for an overview of contacts without temporal focus (e.g., finding a ``super-spreader'').

The \textit{free text feedback} was positive:
``Interesting and well thought-out visualization. I was very enthusiastic about the presentation of this visualization. The visualization of such complex data is not-trivial.''
Participants found some \ch{training} was required, but after understanding it, they found it very useful: ``I find the \storyline really good. Of course, one has to get used to it first and learn to use it first.'' One expert suggested broader applicability ``I would strongly recommend this visualization to be shown to public health authority (Gesundheitsamt).
It could be `fit for purpose'''.

The participants found that scalability in terms of the number of patients posed a significant challenge.
They recommended a two-staged process with filtering through the contact network and then show the filtered data in the \storyline.
Four of the twenty-five participants wanted to change the colors  (e.g., showing each patient in a different color, using more prominent colors for the three areas (home, hospital, temporary home)).
One participant requested further differentiation between carrier and diseased, as it is important for certain pathogens. A number of future extensions were suggested that mainly require additional data.  We discuss them in \autoref{sc:discussion}.

\section{Discussion and Future Work}
\label{sc:discussion}

\begin{figure}[tb]
    \centering
    \begin{tabular}{|c|c|c|c|}
    \hline 
        Layout & Edge crossing & Runtime (ms) & Pre-processing (ms) \\
        \hline 
        Our & 25$^a$ (37$^b$) & 197 & 18  \\
        \cite{liu2013storyflow} & 14 & 160 & $N/A$ \\
        \hline
    \end{tabular}
    \textcolor{gray}{
 \setlength{\fboxsep}{0pt}%
\setlength{\fboxrule}{0.5pt}%
\fbox{\includegraphics[width=\linewidth,height=2.5cm,trim=0 5 0 40 ,clip]{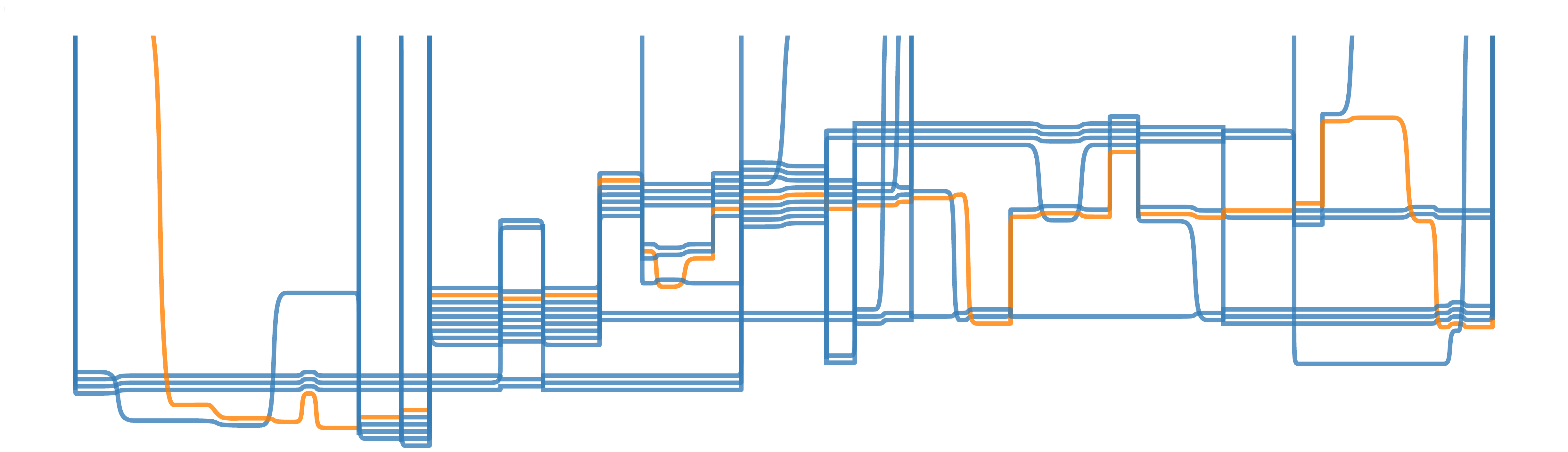}}}
    \caption{Layout quality using Matrix dataset~\cite{tanahashi2013movie}. ($^a$) comparable to Storyflow~\cite{liu2013storyflow}, ($^b$) - including start/end lines.} 
    \label{fig:matrix}
\end{figure}

The \storyline layout was specifically designed for our tasks rather than to optimize edge crossing and wiggles. Still, the approach performs well on the benchmark Matrix dataset, with 14 entities and 90 transfers~\cite{tanahashi2013movie} (see \autoref{fig:matrix}). We used a laptop PC with an Intel i7-9750H CPU (2.6 GHz) and 32GB memory. Pre-processing lasts 18~ms, the initialization stabilized after 30 iterations (62 ms) and force-directed layout after 10 iterations (135 ms). The layout of use case with 27 entities and 704 transfers and the same number of iterations lasted 2517ms in total (190ms + 1265ms + 1062ms). The computation of tracking interaction for all critical patients lasted 37ms in total.

Our data sets pose visual scalability challenges in terms of the time spans of the data sets, number of patients, number of locations, number of patient transfers, and the number of screenings and tests. Interaction is leveraged to allow the approach to scale to larger data sets, combining several views and automatically highlighting transmission pathway events. They allow to explore pathways of up to hundred patients (see Annex). \ch{Explicitly visualizing the number of forward and backward connections of a patient would help the system scale when tracing contacts.}  Other methods for increasing the scalability of our approach should be investigated.

The visual design of \storyline relies heavily on colors to encode infection status and wards, limiting the number of wards that can be represented. \ch{Automatic detection of suspicious wards (i.e., a sudden increase in the number of cases) would help with this scalability along with forms of automated support.}  In order to scale a larger number of wards, interactive highlighting could help and hulls could improve space efficiency~\cite{tanahashi2012design}. The expert feedback indicated a need for enabling annotations to the visualizations with genome data or measures taken during the outbreak, which could be implemented. 

The data collection methodology and privacy considerations pose limitations on what can be analyzed. A core challenge is localizing contacts accurately, which is currently handled via electronic health records. More accurate data on patient mobility through the hospital, locations of rooms and beds, as well as information where procedures occur (e.g., surgery, endoscopy, radiography) could be helpful for certain visual analytics tasks.  Technologies, such as RFID, could be used, but would incur significant privacy considerations. Also, current records do not confirm when a patient has been recovered. Therefore, patients remain infected for the remainder of the data set, which is often not the case. \ch{These data need to be extracted from the current raw data by new algorithms. These algorithms have to be adjusted to the ward, the patient-type, the type of pathogen and the tests.} 
Additional features, like the comparison and matching of antimicrobial profiles simultaneous with the integration of genomics results of the pathogens would help the infection control expert automatically cluster patients according to strains to better identify transmission pathways. \ch{We used an external tool to perform genome analysis to verify the suggested transmission routes. The genome analysis visualization is not part of our system. However, since laboratories have access to near real-time genome analysis, this data can be extracted and used for matching specific pathogens to help verify the transmission pathway hypotheses.  Moreover, the visual interface could be extended to support provenance of analysis and interaction.}
In the future, it would be interesting to encode the mode of transmissions (e.g., airborne or contact) or other potential risk factors (e.g., diabetes, immunotherapy). 

Our approach was designed and implemented for transmission pathway reconstruction in hospitals. \ch{A modified version of the system has now been deployed in several hospitals for analyzing COVID-19 hospital-associated transmissions (see \href{https://www.youtube.com/watch?v=HAsb3dnUKyI&feature=youtu.be}{video in German at https://youtu.be/HAsb3dnUKyI}).} More generally, our approach could be used to visually analyze contact tracing graphs of localized outbreak clusters and in closed environments, such as cruise ships and buildings. \ch{The approach could be applicable to different definitions of spread, such as financial or information contagion. The tracing interaction could help with this reconstruction, but it needs to be adapted to the specifics of the application. For example for information spread, the model would take the beliefs about the information, i.e., whether a person supports or opposes the information~\cite{6785909} into account.} 

%




\section{Conclusion}
\label{sc:concl}

In this paper, we have presented a visual analytics system for visualizing transmission events and outbreaks in hospitals.  The approach focuses on methods for reconstructing the transmission pathways for a certain pathogen back to a \textit{patient zero}.  This visual analytics approach was developed through iterative, user-centered design with seven German Hospitals and Infection Control Institutes or Units.  As part of the design process, storyline visualizations were adapted to visualize temporal contact networks.  Our approach was applied to a real outbreak of \klpn in a large German hospital. Infection control experts were able to effectively unravel the outbreak and reveal two distinct transmission pathways. Tracing back the initial source, \textit{patient zero} was conducted in a faster and in a more comprehensive way when compared to existing methods. \ch{In our final qualitative user study of twenty-five users,} we found significant value in the approach when making sense of outbreaks in hospitals.

\ch{\paragraph*{Acknowledgments}
We would like to thank all infection control experts involved, in particular A.~Wulff, P.~Biermann, S.~Rey, C.~Baier and M.~Kaase, for their helpful feedback. This work was supported by the German Federal Ministry of Education
and Research (BMBF) within the Medical Informatics Initiative (01ZZ1802B/HiGHmed).}

\bibliographystyle{abbrv-doi}
\bibliography{biblio}
\pagebreak
\appendix
\renewcommand\thefigure{\thesection\arabic{figure}}    
\setcounter{figure}{0}

\section{Annex:  Genome Analysis}

To provide further support and verification for the hypothesized transmission pathways, the experts also used genome information of the pathogen. Comparing genomes of bacterial strains adds a further evidence to the analysis of outbreaks. This tool does not belong to the described system but represents a standard tool to describe the relatedness of pathogens collected from patients. 

In brief, the DNA of each \klpnShort was extracted (using Qiagen DNeasy Blood \& Tissue) and sequenced using the illumina MiSeq system (sequences available from the authors on request).

\begin{figure}[tbph]
\centering
    \textcolor{gray}{
 \setlength{\fboxsep}{0pt}%
\setlength{\fboxrule}{0.5pt}%
\fbox{ \includegraphics[trim=0 0 0 0,clip,width=0.9\linewidth]{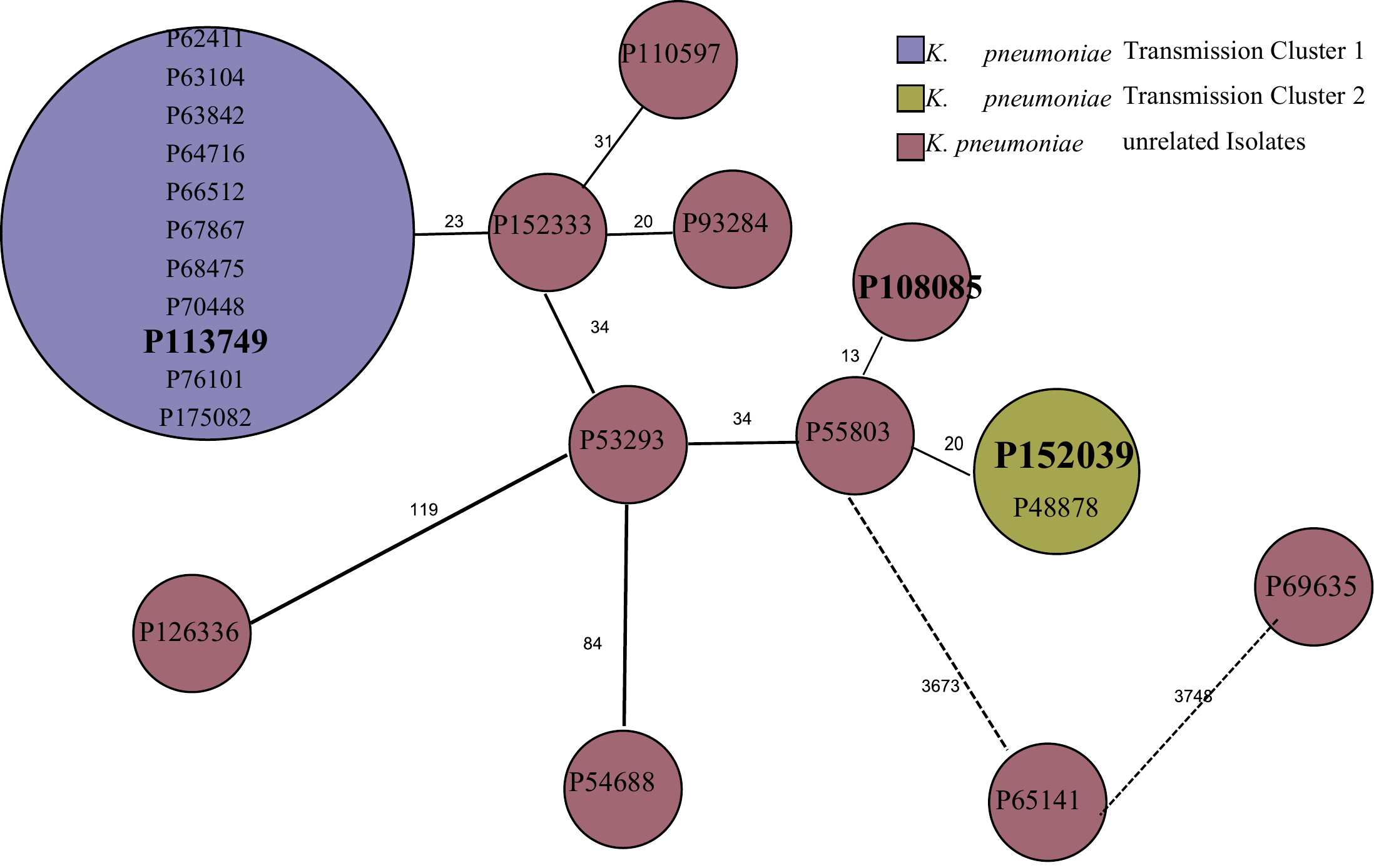}}}
     \caption{The genome analysis of patients infected by \klpnShort bacteria strain. The numbers on the line indicate the number of different alleles in the wgMLST scheme. Bacterial strains were regarded as identical/similar when less than five alleles differed. The visualization is a minimum spanning tree and was done with \href{https://www.applied-maths.com/bionumerics}{BioNumerics}.} 
    \label{fig:UCgenome}
\end{figure}

In total, the genomes of 23 bacterial isolates were analyzed and compared using the whole genome MultiLocus Sequence Typing (wgMLST) scheme for \klpnShort~\cite{bialek2014Kleb} (BioNumerics v7.6 created by Applied Maths NV). The results represent the genome relation of the bacteria strains in a minimum spanning tree. Identical or very similar genomes (less than 5 allelic differences) were regarded as the same strain and support a direct transmission. The genome analysis confirms that eleven patients on the subwards S279 and S295 were infected by one \klpnShort bacteria strain, including the potential \textit{patient zero} candidate P113749. The bacterial strains of the patients show no differences in the wgMLST profile. 
Interestingly, the two other potential \textit{patients zero} P108085 and P152039 reveal substantial differences to this cluster of patients (see \autoref{fig:UCgenome}). Based on the data, the expert assigns patient P113749 as \textit{patient zero} of this particular outbreak on the subwards. The genome analysis reveals patient P1520395 as a further source of infection for at least one more patient on the considered wards. Thus, the two transmission pathways occurred simultaneously.  
\newpage


\end{document}